\documentclass[a4paper,onecolumn,11pt,accepted=2024-09-12]{quantumarticle}
\pdfoutput=1

\PassOptionsToPackage{compress}{natbib}
\usepackage[numbers]{natbib}

\usepackage[utf8]{inputenc}
\usepackage[english]{babel}
\usepackage[T1]{fontenc}
\usepackage{amsmath}
\usepackage{amssymb}
\usepackage{hyperref}
\usepackage{verbatim}
\usepackage{color}
\usepackage{dcolumn}
\usepackage{bm}
\usepackage{dutchcal}

\usepackage{tikz}
\usepackage{lipsum}

\makeatletter
\DeclareFontFamily{OMX}{MnSymbolE}{}
\DeclareSymbolFont{MnLargeSymbols}{OMX}{MnSymbolE}{m}{n}
\SetSymbolFont{MnLargeSymbols}{bold}{OMX}{MnSymbolE}{b}{n}
\DeclareFontShape{OMX}{MnSymbolE}{m}{n}{
    <-6>  MnSymbolE5
   <6-7>  MnSymbolE6
   <7-8>  MnSymbolE7
   <8-9>  MnSymbolE8
   <9-10> MnSymbolE9
  <10-12> MnSymbolE10
  <12->   MnSymbolE12
}{}
\DeclareFontShape{OMX}{MnSymbolE}{b}{n}{
    <-6>  MnSymbolE-Bold5
   <6-7>  MnSymbolE-Bold6
   <7-8>  MnSymbolE-Bold7
   <8-9>  MnSymbolE-Bold8
   <9-10> MnSymbolE-Bold9
  <10-12> MnSymbolE-Bold10
  <12->   MnSymbolE-Bold12
}{}

\let\llangle\@undefined
\let\rrangle\@undefined
\DeclareMathDelimiter{\llangle}{\mathopen}%
                     {MnLargeSymbols}{'164}{MnLargeSymbols}{'164}
\DeclareMathDelimiter{\rrangle}{\mathclose}%
                     {MnLargeSymbols}{'171}{MnLargeSymbols}{'171}
\makeatother

\newcommand{\Lin}[1]{\mathsf{L}_{#1}}
\newcommand{\Hin}[1]{\mathsf{H}_{#1}}
\newcommand{\Din}[1]{\mathsf{D}_{#1}}
\newcommand{\Kin}[1]{\mathsf{K}[#1]}
\newcommand{\betatx}[1]{\beta^{\text{#1}}}
\newcommand{\cl}{\text{cl}}
\newcommand{\qu}{\text{qu}}

\begin{document}

\title{Stochastic Thermodynamics at the Quantum-Classical Boundary:
A Self-Consistent Framework Based on Adiabatic-Response Theory}

\author{Joshua Eglinton}
\affiliation{School of Physics and Astronomy, University of Nottingham, Nottingham NG7 2RD, United Kingdom}
\affiliation{Centre for the Mathematics and Theoretical Physics of Quantum Non-Equilibrium Systems,
University of Nottingham, Nottingham NG7 2RD, United Kingdom}
\orcid{0000-0001-7055-0807}

\author{Federico Carollo}
\affiliation{Institut f\"ur Theoretische Physik, Eberhard Karls Universit\"at T\"ubingen,
Auf der Morgenstelle 14, 72076 T\"ubingen, Germany}
\orcid{0000-0002-6961-7143}

\author{Igor Lesanovsky}
\affiliation{School of Physics and Astronomy, University of Nottingham, Nottingham NG7 2RD, United Kingdom}
\affiliation{Centre for the Mathematics and Theoretical Physics of Quantum Non-Equilibrium Systems,
University of Nottingham, Nottingham NG7 2RD, United Kingdom}
\orcid{0000-0001-9660-9467}
\affiliation{Institut f\"ur Theoretische Physik, Eberhard Karls Universit\"at T\"ubingen,
Auf der Morgenstelle 14, 72076 T\"ubingen, Germany}

\author{Kay Brandner}
\affiliation{School of Physics and Astronomy, University of Nottingham, Nottingham NG7 2RD, United Kingdom}
\affiliation{Centre for the Mathematics and Theoretical Physics of Quantum Non-Equilibrium Systems,
University of Nottingham, Nottingham NG7 2RD, United Kingdom}
\orcid{0000-0002-4425-8252}
\email{kay.brandner@nottingham.ac.uk}

\begin{abstract}
Microscopic thermal machines promise to play an important role in future quantum technologies. Making such devices widely applicable will require effective strategies to channel their output into easily accessible storage systems like classical degrees of freedom. 
Here, we develop a self-consistent theoretical framework that makes it possible to model such quantum-classical hybrid devices in a thermodynamically consistent manner. 
Our approach is based on the assumption that the quantum part of the device is subject to strong decoherence and dissipation induced by a thermal reservoir. 
Due to the ensuing separation of time scales between slowly evolving classical and fast relaxing quantum degrees of freedom, the dynamics of the hybrid system can be described by means of adiabatic-response theory. 
We show that, upon including fluctuations in a minimally consistent way, the resulting equations of motion can be equipped with a first and second law, both on the ensemble level and on the level of individual trajectories of the classical part of the system, where thermodynamic quantities like heat and work become stochastic variables. 
As an application of our theory, we work out a physically transparent model of a quantum-classical hybrid engine, whose working system consists of a chain of Rydberg atoms, which is confined in an optical cavity and driven by periodic temperature variations. 
We demonstrate through numerical simulations that the engine can sustain periodic oscillations of a movable mirror, which acts as a classical load, against external friction and extract the full distributions of input heat and output work. 
By making the statistics of thermodynamic processes in quantum-classical hybrid systems accessible without the need to further specify a measurement protocol, our work contributes to bridging the long-standing gap between classical and quantum stochastic thermodynamics.
\end{abstract}

\maketitle

\newpage
\section{Introduction}

Thermal machines have played a central role in the development of classical thermodynamics, whose early pioneers sought to uncover the fundamental principles that determine the performance of devices like heat engines or refrigerators. 
Almost two centuries later, understanding the working mechanisms of microscopic thermal machines, which operate on atomistic length and energy scales, and unlocking their potential for future applications have become major driving forces of quantum thermodynamics \cite{Kosloff2014,Vinjanampathy2016,Benenti2017,Mukherjee2021,Myers2022}.
The last decade has seen remarkable progress in this direction. On the theory side, a whole spectrum of new concepts has emerged, for example to exploit collective effects in quantum many-body systems for power generation and cooling \cite{Hardal2015,Uzdin2016,Jaramillo2016,Ma2017,Niedenzu2018,Li2018,Chen2019,Latune2019,
Myres2020,Keller2020,Watanabe2020,Carollo2020,Carollo2020b,Fogarty2021,Myres2021,Kloc2021,
Myers2022b,Li2022,Souza2022,Paulino2023,Macovei2022,Kolisnyk2023,Yadin2023,Marzolino2024,
Eglinton2023}, 
to identify optimal control strategies for quantum thermodynamic cycles \cite{Eglinton2023,Campo2014,Abiuso2020,Miller2020,Bhandari2020,Brandner2020,Alonso2022,
Eglinton2022,Menczel2019,Hartmann2020,Pancotti2020,Xu2021,Khait2022,Erdman2022,
Das2023,Deng2024}, 
to reduce fluctuations in quantum thermal machines
\cite{Souza2022,Pietzonka2018,Campisi2015,Saryal2021,Liu2021,Miller2021,Rignon2021,Gerry2022,
Xiao2023}, 
or to channel their output into externally accessible storage systems
\cite{Levy2016,Binder2015,Roulet2017,Seah2018,Van2020,Martins2023,Leitch2024,Lindenfels2019}.
On the experimental side, microscopic thermal machines have been realized on a whole variety of different platforms including single atoms and ions \cite{Lindenfels2019,Abah2012,Robnagel2016}, ultracold atomic gases
\cite{Bouton2021,Koch2023}, 
semiconductor nano-structures \cite{Josefsson2018}, 
superconducting devices \cite{Ronzani2018},
photonic systems \cite{Kim2022} and nuclear spins 
\cite{Peterson2019,Lisboa2022}. 

These developments have transformed our conceptual understanding of both the very nature of thermal machines and the criteria by which their performance should be assessed. Besides traditional figures of merit such as power and efficiency, the list of desirable properties now includes scalability, i.e., the ability to increase outputs by enlarging working systems, constancy, i.e., resilience against thermal and quantum fluctuations, and exploitability, i.e., the possibility to access and store generated outputs.
Addressing as many as possible of these criteria at the same time is a challenging endeavour, which quickly leads to fundamental problems. 
Quantities like work and heat, for example, can in general not be measured in quantum systems without altering their state or even their internal energy
\cite{Talkner2007,Esposito2009,Campisi2011,Talkner2016,Perarnau2017,Niedenzu2019,Jacob2022,
Jacob2023}. 
As a result, the actual output of quantum thermal machines can, even in theory, only be determined in the context of specific measurement protocols, which can be hard to implement in practice and do not necessarily reflect realistic working conditions \cite{Talkner2007,Esposito2009,Campisi2011,Talkner2016,Perarnau2017,Niedenzu2019,
Jacob2022,Jacob2023,Strasberg2019}.
  
An elegant means of circumventing this problem would be to channel the output of quantum thermal machines into classical systems, which, at least in principle, can be monitored without perturbing their state. 
However, this idea immediately leads to the question of how to describe the dynamics of hybrid systems with both classical and quantum degrees of freedom, which, despite long-standing efforts has not been fully settled so far
\cite{Sherry1978,Sherry1979,Boucher1988,Kuo1993,Anderson1995,Prezhdo1997,Diosi1998,Caro1999,
Diosi2000,Kapral2006,Barcelo2012,Salcedo2012,Diosi2014,Kapral2015,Burghardt2021,Oppenheim2022,
Oppenheim2023,Layton2023,Lajos2023,Manjarres2024,Alessio2014}.

Here, we seek to address this problem from the perspective of quantum thermodynamics. 
Our approach is based on the assumption that the quantum degrees of freedom are coupled to a thermal reservoir, which induces strong damping and decoherence on a fast time scale. 
As a result, the state of the quantum subsystem remains close to thermal equilibrium, even under the influence of classical degrees of freedom, which can typically be assumed to evolve on much slower time scales than the quantum ones. 
At the same time, any measurement-induced backaction from the classical on the quantum subsystem becomes negligible compared to reservoir-induced decoherence. 
Under these conditions, the evolution of the quantum degrees of freedom can be described by means of adiabatic perturbation theory, a general method to describe systems with well separated time scales, which has been used before to model quantum-classical hybrid dynamics
\cite{Zhang2006,Thomas2012,Alessio2014} 
and has recently been deployed very successfully in quantum thermodynamics \cite{Eglinton2023,Abiuso2020,Miller2020,Bhandari2020,Brandner2020,Alonso2022,Eglinton2022,
Miller2021,Thomas2012,Cavina2017,Miller2019}. 
The effective force that is exerted by the quantum subsystem on the classical one, which can in principle be driven arbitrarily far from equilibrium, can then be inferred from energy conservation. 
Upon including fluctuations in a minimal thermodynamically consistent way, it thus becomes possible to construct a self-consistent framework that describes the dynamics and thermodynamics of quantum-classical hybrid systems in terms of stochastic equations of motion and a first and a second law.
Furthermore, since the classical degrees of freedom follow well-defined observable trajectories, a concept that does in general not exist for quantum systems, these laws can be formulated on the level of ensembles and for individual realizations of the dynamics. 
Thermodynamic processes can therefore be analyzed in a statistical manner, where quantities like work and heat become stochastic variables, whose full distributions can, at least in principle, be inferred by observing only the classical degrees of freedom.
As a result, many of the established concepts of classical stochastic thermodynamics \cite{Seifert2008,Seifert2012,Seifert2018} can be directly applied to quantum-classical hybrid systems. In the context of thermal machines, for instance, output fluctuations can be determined by observing the trajectories of a classical load. 
Constancy, as a third figure of merit besides power and efficiency \cite{Pietzonka2018}, thus becomes a directly accessible parameter without the need of further specifying a measurement protocol. 

\begin{figure}[t]
    \centering
    \includegraphics[width=0.6\textwidth]{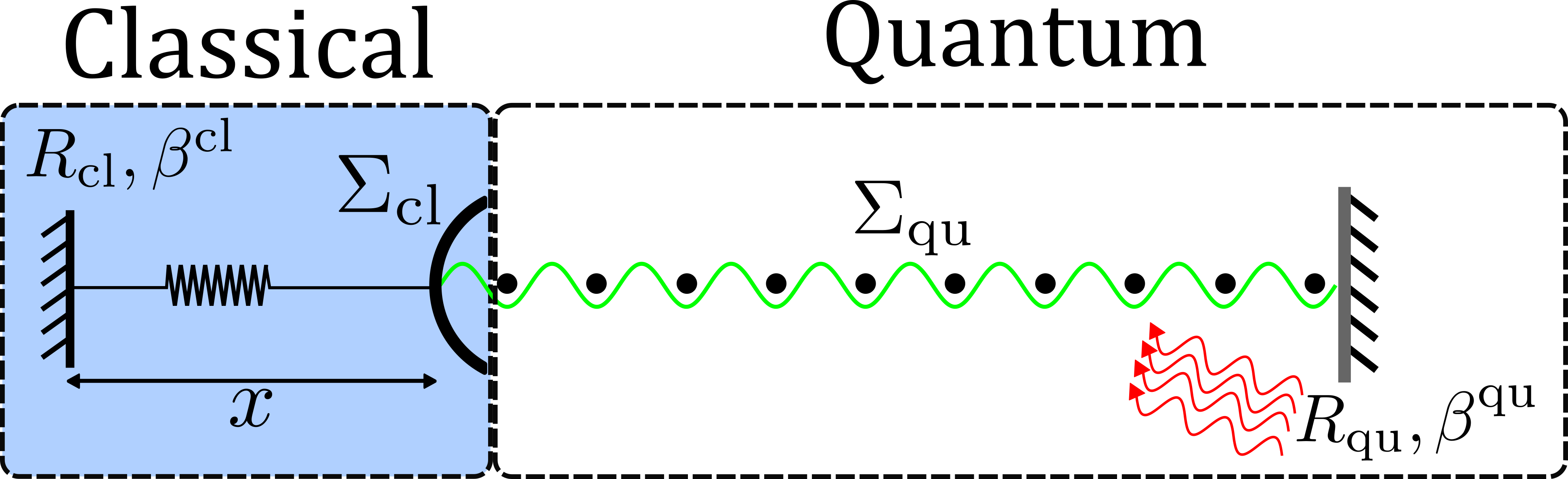}
    	\caption{Quantum-classical hybrid system. 
    		The classical and the quantum subsystem, $\Sigma_{\cl}$ and $\Sigma_{\qu}$, 
    		respectively consist of a movable mirror and a chain of Rydberg atoms inside an 
    		optical cavity. 
    		Both subsystems are coupled to thermal reservoirs, $R_{\cl}$ and $R_{\qu}$, 
    		with inverse temperatures $\betatx{cl}$ and $\betatx{qu}$.}
    	\label{fig:schem}
\end{figure}

From a practical perspective, it is important to identify suitable experimental platforms, on which quantum-classical hybrid machines can be implemented and theoretical predictions can be tested. 
Such experiments can be expected to be challenging, since they require matching the typically vastly different energy scales of classical and quantum objects, while maintaining at least some non-classical properties of the quantum degrees of freedom, which can quickly be degraded by the influence of the classical ones, see Sec.~\ref{Sec:Perspectives} for a quantitative discussion. 
Nonetheless, opto-mechanical systems, which can be realized in a variety of different ways \cite{Aspelmeyer2014}, for example, with cold atoms \cite{Murch2008,Brennecke2008,Schleier2011}, photonic crystals \cite{Gavartin2011}, or superconductors coupled to micro-mechanical membranes \cite{Youssefi2022}, provide a natural starting point.
The mechanical degrees of freedom of such systems, like movable mirrors, can, at least in the weak-coupling regime, often be treated classically, while optical ones, like the electromagnetic field modes of a cavity, retain their quantum properties. 
Hybrid models of this kind can be derived from a full quantum-mechanical description through mean-field type approximations, where quantum fluctuations in the mechanical subsystem are neglected; that is, the corresponding operators are replaced with their average values.
While it is yet to be rigorously established under what conditions and by what mechanisms quantum-classical hybrid dynamics emerge from full quantum models, this semi-phenomenological approach proved very useful to describe a whole range of opto-mechancial experiments \cite{Aspelmeyer2014}.

In the second part of this article, we use this platform to develop a physically transparent model of a quantum-classical hybrid engine, whose classical degree of freedom consists of a movable mirror, which, together with a second fixed mirror, forms an optical cavity.
The quantum part of the engine is formed by a long chain of Rydberg atoms, whose large internal energy scale makes them natural candidates for quantum systems that can plausibly exert an observable effect on a classical object \cite{Adams2019}. 
Similar to a recent experiment, where a quantum heat engine was realized on the platform of super-radiant atoms \cite{Kim2022}, the interaction between the two subsystems is mediated by the radiation pressure inside the cavity. 
Upon being provided with input from a noisy laser, which mimics a thermal reservoir, the machine can sustain periodic oscillations of the movable mirror, which plays the role of a classical load, against classical friction, see Fig.~\ref{fig:schem}. 
As we show by means of numerical simulations, our framework makes it possible to obtain the full distributions of output work and effective input heat for this model, which illustrates the potential of our theory to open a new perspective on the stochastic thermodynamics of quantum systems. 

\section{\label{sec:Therm fw} Dynamical framework}

We first consider a generic hybrid system that consists of a quantum subsystem $\Sigma_{\qu}$ and a classical subsystem $\Sigma_{\cl}$ in contact with their respective reservoirs $R_{\qu}$ and $R_{\cl}$ whose inverse temperatures are $\betatx{qu}$ and $\betatx{cl}$, see Fig.~\ref{fig:schem}. 
The two subsystems are mechanically coupled so that the Hamiltonian of $\Sigma_{\qu}$ depends on the state of $\Sigma_{\cl}$. 
In this section we develop a general framework to describe the non-equilibrium dynamics of this setup. Our approach is based on the assumption that the characteristic timescales of the classical and quantum subsystems, that is the observational timescales, $\tau^{\Sigma_{\cl}}$ and $\tau^{\Sigma_{\qu}}$, on which the state of $\Sigma_{\cl}$ and $\Sigma_{\qu}$ change significantly, are well separated. 
As we will show in the next section, the resulting dynamical framework is fully consistent with the laws of thermodynamics. 

\subsection{Quantum subsystem}

We initially consider the state of the classical subsystem $\Sigma_{\cl}$ to be frozen. 
The quantum subsystem can then be regarded as an open system, whose Hamiltonian $H_{x}^{\qu}$ depends on a fixed external parameter $x$, for instance the position of a movable mirror of an optical cavity. 
We take the Hilbert space of the quantum subsystem to be finite-dimensional throughout. 
The state of $\Sigma_{\qu}$ is described by the density matrix $\rho$, which evolves in time according to the quantum master equation
\begin{equation}
\label{eq: ME qu}
    \dot{\rho}_t = \Lin{x}\rho_t.
\end{equation}
In the weak-coupling limit, on which we focus here, the generator $\Lin{x}$ has the generic form
\begin{equation}
\label{eq: ME}
    \Lin{x} \circ=\Hin{x} \circ+ \Din{x} \circ
\end{equation}
with the two superoperators
\begin{subequations}\label{eq: GenDef}
    \begin{align}
    \Hin{x}\circ&=-\frac{i}{\hbar}\big[H_{x}^{\qu},\circ\big]-\frac{i}{\hbar}\big[H_{x}^{\text{LS}},\circ\big], 
    \\
    \Din{x}\circ&=\sum_{j}\Big\{\gamma_{x}^{j-} \ \Kin{L_x^j}\circ+\  \gamma_{x}^{j+} \ \Kin{L_x^{j\dagger}}\circ\Big\},
    \end{align}
\end{subequations}
where we have introduced the short-hand notation
\begin{equation}
\label{eq: kin}
    \Kin{Y}\circ = Y\circ Y^{\dagger} - \frac{1}{2} Y^{\dagger}Y\circ - \frac{1}{2}\circ Y^{\dagger}Y.
\end{equation}
The Lamb shift $H_{x}^{\text{LS}}$ and the Lindblad operators $L_x$ and $L_x^{\dagger}$ account for the influence of the reservoir $R_{\qu}$ on $\Sigma_{\qu}$ and have the following properties. 
The Lamb shift commutes with the Hamiltonian of the quantum subsystem $H^{\qu}_x$ and therefore only induces corrections to its Bohr frequencies. 
The Lindblad operators induce jumps between energy levels of $H^{\qu}_x$, during which the system absorbs or emits the energy $\hbar\nu_{x}^{j}\geq0$. 
That is, we have
\begin{subequations}
\label{eq: DB cond}
    \begin{align}
    &\big[H_{x}^{\qu},H_{x}^{\text{LS}}\big]=0,
        \\
        &\big[H_{x}^{\qu},L_{x}^{j}\big]=-\hbar\nu_{x}^{j}L_{x}^{j},\qquad\big[H_{x}^{\qu},L_{x}^{j\dagger}\big]=\hbar\nu_{x}^{j}L_{x}^{j\dagger}.
    \end{align}
\end{subequations}
Micro-reversibility further requires the dissipation rates $\gamma_{x}^{j+}$ and $\gamma_{x}^{j-}$ to obey the detailed balance condition
\begin{equation}
    \label{eq: micro rev}\gamma_{x}^{j+}/\gamma_{x}^{j-}=\exp\big[-\betatx{qu}\hbar\nu_{x}^{j}\big],
\end{equation}
which is required for thermodynamic consistency. 
We note that the conditions \eqref{eq: DB cond} and \eqref{eq: micro rev} require that the Lamb shift, the Lindblad operators and the dissipation rates are dependent on $x$, since $H^{\qu}_x$ depends on $x$. 
As long as this parameter is fixed, the stationary solution of the master equation \eqref{eq: ME} is given by the Gibbs state
\begin{equation}
\label{eq: Gibss state}
    \varrho_{x}^{\qu}\equiv\exp\Big[-\betatx{qu}\big(H_{x}^{\qu}-\mathcal{F}_{x}\big)\Big],
\end{equation}
where $\mathcal{F}_{x}$ is the free energy of the quantum subsystem.

\subsection{Classical subsystem}

We now identify the parameter $x$ with the position of the classical subsystem $\Sigma_{\cl}$, whose full state is described by the phase space vector $\mathbf{z}\equiv(x,p)$ and whose time evolution is governed by the Langevin equations
\begin{subequations}
    \begin{align}
        \dot{x}_t&=p_t/m,
        \\
        \dot{p}_t&=-U_{x_t}'+f_{\mathbf{z}_t}^{\qu}-\zeta^{\cl}p_t+\xi^{\cl}_t.
    \end{align}
\end{subequations}
Here, $m$ denotes the mass of the classical degree of freedom, $U_{x}$ is an external potential and primes denote derivatives with respect to $x$. 
Furthermore, $f_{\mathbf{z}}^{\qu}$ denotes the force that the quantum subsystem exerts on the classical one. 
The Gaussian stochastic force $\xi_{t}^{\cl}$, which together with the friction constant $\zeta^{\cl}$ accounts for the influence of the classical reservoir $R_{\cl}$, obeys
\begin{equation}
\label{eq: cl xi}
    \langle \xi_{t}^{\cl}\rangle= 0\quad\text{and}\quad\langle \xi_{t}^{\cl} \xi_{t'}^{\cl}\rangle\equiv 2D^{\cl}\delta_{t-t'},
\end{equation}
where $\delta_{t}$ denotes Dirac delta.
The diffusion constant $D^{\cl}$ is thereby related to the friction constant $\zeta^{\cl}$ through the fluctuation-dissipation theorem \cite{Kubo2012}
\begin{equation}
\label{eq: cl fluc dis}
    D^{\cl}=\zeta^{\cl}m/\betatx{cl}.
\end{equation}

\subsection{Quantum coupling force}

To determine the coupling force $f_{\mathbf{z}}^{\qu}$, we assume that $x_t$ varies slowly on the relaxation timescale of the quantum subsystem $\Sigma_{\qu}$. 
Under this condition, the time evolution of $\Sigma_{\qu}$ can be described in terms of an adiabatic quantum master equation \cite{Cavina2017}, which can be obtained from Eq.~\eqref{eq: ME qu} by replacing $x\rightarrow x_t$. 
The solution of this equation can be found by means of adiabatic perturbation theory. 
Here we use the second-order ansatz
\begin{equation}
\label{eq: rho ad resp}
    \rho_t=\varrho_{x_t}^{\qu}+\rho_{x_t}^{(x)}p_t+\rho_{x_t}^{(2x)}p_t^2+\rho_{x_t}^{(p)}\dot{p}_t
\end{equation}
for the state of $\Sigma_{\qu}$ \cite{Cavina2017,Miller2019,Brandner2020}, which will turn out to be a minimal consistent choice. 
We note that, in the quasi-static limit, we have $\rho_t \rightarrow\varrho^{\qu}_{x_t}$. 
That is, the system follows its instantaneous equilibrium state, which depends only on $x$ but not on $p$. 
Therefore, the term proportional to $\dot{p}_t$ in Eq.~\eqref{eq: rho ad resp} is entirely of second order with respect to the adiabatic perturbation theory and no higher-order derivatives of $p_t$ or cross terms between $p_t$ and $\dot{p}_t$ appear.
Inserting Eq.~\eqref{eq: rho ad resp} into the adiabatic master equation yields
\begin{equation}
\label{eq: 2nd order rho}
    \rho_{x}^{(x)}=\frac{1}{m}\Lin{x}^{-1}{\varrho_{x}^{\qu}}',\qquad
        \rho_{x}^{(2x)}=\frac{1}{m}\Lin{x}^{-1}{\rho_{x}^{(x)}}',\qquad
        \rho_{x}^{(p)}=\Lin{x}^{-1}{\rho_{x}^{(x)}},
\end{equation}
where we have used that generator $\mathsf{L}_x$ defined in Eqs.~\eqref{eq: ME} and \eqref{eq: GenDef} is independent of $p$.
Furthermore, we assume that $\varrho_{x}^{\qu}$ is the only stationary state of the frozen generator $\Lin{x}$ such that the super-operator inverse in Eqs.~\eqref{eq: 2nd order rho} is well defined \footnote{Since ${\varrho_{x}^{\qu}}{'}$ and ${\rho_{x}^{(x)}}{'}$ have vanishing trace, these operators are orthogonal to $\varrho_{x}^{\qu}$ with respect to the scalar product $\langle X|Y\rangle=\int_{0}^{1} d\lambda \ \text{Tr}\big[(\varrho_{x}^{\qu})^{1-\lambda}X^{\dagger}(\varrho_{x}^{\qu})^{\lambda}Y\big]$, which renders the super operator $\Lin{x}$ self adjoint.}. 
The adiabatic-response approach is justified if $x$ changes slowly on the relaxation timescale $\tau^{\Sigma_{\qu}}$ of the quantum subsystem. 
That is, we assume the timescale $\tau^{\Sigma_{\cl}}$ is large compared with the quantum relaxation timescale set by the rates $\gamma_{x}^{j+}$ and $\gamma_{x}^{j-}$.
In addition, by using an adiabatic master equation, we implicitly assume that the characteristic time scale of the frozen unitary dynamics of the quantum subsystem, which is set by the Bohr frequencies of the Hamiltonian $H^{\qu}_x$, is short compared to $\tau^{\Sigma_{\cl}}$, so that coherent oscillations in the interaction picture with respect to $H^{\qu}_x$ average out before the position $x$ of the classical degree of freedom changes significantly, for further details, see, for instance, the Supplemental Material of Ref.~\cite{Brandner2020}.

The systematic part of the coupling force $f_{\mathbf{z}}^{\qu}$ reads
\begin{equation}
\label{eq: F=-H}
    \langle f_{\mathbf{z}}^{\qu}\rangle=\text{Tr}\big[F_{x}\rho\big]\quad\text{with}\quad F_{x}\equiv -{H_{x}^{\qu}}'
\end{equation}
playing the role of a quantum force operator.
This definition is motivated by the observation that, if the internal energy of the quantum systems is identified as $E^{\qu} = \text{Tr} [H_x \rho]$, an infinitesimal change of the $x$ at fixed $\rho$ leads to the energy increment 
\begin{equation}
 dE^{\qu} = \text{Tr}[H^{\qu\prime}_x \rho] dx. 
\end{equation}
In analogy with classical mechanics, the right-hand side of this equation can be interpreted as the product of a force and an infinitesimal displacement, which immediately leads to the formula \eqref{eq: F=-H} for the coupling force, if the sign of the latter is fixed so that a positive force performs work on the classical subsystem \cite{Alessio2014}.
In adiabatic response, this expression must be evaluated using the first-order adiabatic expansion of $\rho_t$, so that it can be consistently interpreted as a force acting on the classical degree of freedom, which depends only on its position and velocity, but not on acceleration. 
We thus obtain
\begin{equation}
    \langle f_{\mathbf{z}}^{\qu}\rangle = \text{Tr}\big[F_{x}\varrho_{x}^{\qu}\big] + \text{Tr}\big[F_{x}\rho_{x}^{(x)}\big]p=-\mathcal{F}_{x}'-\zeta_{x}^{\qu}p,
\end{equation}
with the quantum friction constant
\begin{equation}
    \zeta_{x}^{\qu}=-\text{Tr}\big[F_{x}\rho^{(x)}_x\big].
\end{equation}
The resulting quantum friction force has to be counterbalanced by the fluctuating force $\xi_{x,t}^{\qu}=f_{\mathbf{z},t}^{\qu}-\langle f_{\mathbf{z},t}^{\qu}\rangle$, which accounts for both thermal and quantum fluctuations in the quantum subsystem. 
Under our overarching assumption that the quantum degrees of freedom evolve fast on the observational time scale of the classical ones, $\tau^{\Sigma_{\cl}}\ll \tau^{\Sigma_{\qu}}$, this force can be approximately described as Gaussian and uncorrelated with the classical fluctuating force \cite{Bode2011,Thomas2012}. 
Hence, we set
\begin{equation}
\label{eq: qu fluc rel}
    \langle \xi_{x,t}^{\qu}\rangle=\langle\xi_{x,t}^{\qu}\xi_{t'}^{\cl}\rangle=0\quad\text{and}\ \ \ \langle\xi_{x,t}^{\qu}\xi_{x,t'}^{\qu}\rangle =2D_{x}^{\qu}\delta_{t-t'}.
\end{equation}
The quantum diffusion constant $D_{x}^{\qu}$ can now be identified with the equilibrium correlation function
\begin{equation}
\label{eq: Dqu def}
    D_{x}^{\qu} = \int_0^{\infty} dt \ \llangle \delta \hat{F}_{x,t} ;\delta \hat{F}_{x,0}\rrangle_{\varrho^{\qu}}
   =-\llangle\big(\Lin{x}^{\ddagger}\big)^{-1}\delta F_{x};\delta F_{x}\rrangle_{\varrho^{\qu}}\geq0,
\end{equation}
where $\delta F_{x}=F_{x}+\mathcal{F}_{x}'$ and hats indicate Heisenberg-picture operators \cite{Thomas2012}. The super-operator $\Lin{x}^{\ddagger}$ is the adjoint of $\Lin{x}$ with respect to the Hilbert-Schmidt scalar product and the Kubo correlation function that appears in Eq.~\eqref{eq: Dqu def} is defined as \cite{Kubo2012}
\begin{equation}
    \llangle \circ ;\bullet \rrangle_{Y}\equiv\int_{0}^{1} d\lambda \ \text{Tr}\big[\circ^{\dagger}Y^{\lambda}\bullet Y^{1-\lambda}\big].
\end{equation}
We note that, in contrast to a symmetrized correlation function, which would have been a second natural choice for $D^{\qu}_x$, the Kubo correlation function ensures that the fluctuation-dissipation theorem
\begin{equation}
\label{eq: Dq fdr}
    D_{x}^{\qu}=\zeta_{x}^{\qu}m/\betatx{qu}
\end{equation}
holds\footnote{We note that that the symmetrized and Kubo correlation functions are equivalent for coherent mesoscopic conductors, i.e. the systems considered in Refs.~\cite{Bode2011,Thomas2012}.}.
As we will see in the following, this condition is necessary for the system to relax to a global thermal equilibrium state, which is free of dissipative currents, if $\betatx{qu}=\betatx{cl}$. 
The total coupling force between the classical and quantum subsystem is now given by
\begin{equation}
    f_{\mathbf{z},t}^{\qu}=-\mathcal{F}_{x}'-\zeta_{x}^{\qu}p+\xi_{x,t}^{\qu},
\end{equation}
where the fluctuating force $\xi_{x,t}^{\qu}$ is fully characterized by the relations \eqref{eq: qu fluc rel} and \eqref{eq: Dq fdr}. 
We therefore have a complete dynamical model of our quantum-classical hybrid system.

\subsection{Fokker-Planck equation}

The classical probability density $P_{\mathbf{z},t}$ of finding $\Sigma_{\cl}$ in the state $\mathbf{z}=(x,p)$ obeys the Fokker-Planck equation
\begin{equation}
\label{eq: FP eq}
    \dot{P}_{\mathbf{z},t}=-\partial_x j_{\mathbf{z},t}^{x} - \partial_{p}j_{\mathbf{z},t}^{p},
\end{equation}
where the probability currents $j_{\mathbf{z},t}^{x}\equiv j^x_t$ and $j^p_{\mathbf{z},t}\equiv j^p_t$ are defined as
\begin{subequations}\label{eq: prob currents}
    \begin{align}
        j^x_t& \equiv \frac{p}{m}P_t,
        \\
        j^p_t &\equiv -\Big(U_{x}'+\mathcal{F}_{x}'+\big(\zeta^{\cl}+\zeta_{x}^{\qu}\big)p+\big(D^{\cl}+D_{x}^{\qu}\big)\partial_p\Big)P_t
    \end{align}
\end{subequations}
with $P_t\equiv P_{\mathbf{z},t}$. 
Once the solution of Eq.~\eqref{eq: FP eq} for a given initial distribution has been found, the average of an arbitrary observable $A_{\mathbf{z}}$ of the system can be evaluated as
\begin{equation}\label{eq: full av}
    \langle A_{\mathbf{z}}\rangle_t=\int d\mathbf{z} \ \text{Tr}\big[A_{\mathbf{z}}\rho_{\mathbf{z},t}\big].
\end{equation}
The extended density matrix $\rho_{\mathbf{z},t}$, which describes the joint state of $\Sigma_{\qu}$ and $\Sigma_{\cl}$, thereby depends explicitly on the phase-space variables $\mathbf{z}$ and is given by
\begin{equation}
\label{eq: rho ad exp}
    \rho_{\mathbf{z},t}\equiv \varrho_{x}^{\qu}P_t + \rho_{x}^{(x)}pP_t+\rho_{x}^{(2x)}p^{2}P_t+\rho_{x}^{(p)}j^p_t.
\end{equation}
This expression follows by interpreting the density matrix given in Eq.~\eqref{eq: rho ad resp} as describing the state of the quantum subsystem under the condition that the values of the classical phase space variables $x_t$, $p_t$ and $\dot{p}_t$ are known.
Since the adiabatic-response corrections defined in Eq.~\eqref{eq: 2nd order rho} are all traceless, $\rho_{\mathbf{z},t}$ satisfies the normalization conditions
\begin{equation}
    \text{Tr}\big[\rho_{\mathbf{z},t}\big]=P_t\quad\text{and}\ \int d\mathbf{z} \  \text{Tr}\big[\rho_{\mathbf{z},t}\big] =1.
\end{equation}

In equilibrium, i.e., for $\betatx{qu}=\betatx{cl}=\beta$, the stationary solution of Eq.~\eqref{eq: FP eq} is
\begin{equation}
\label{eq: cl P}
    P_{\text{eq}}=\exp{\Big[-\beta\big(H_{\mathbf{z}}^{\cl}+\mathcal{F}_{x}-\mathcal{F}_{\mathbf{z}}^{\cl}\big)\Big]}
\end{equation}
with the Hamiltonian of the classical subsystem
\begin{equation}
    H_{\mathbf{z}}^{\cl}\equiv U_{x} + p^2/2m
\end{equation}
and the classical free energy $\mathcal{F}_{\mathbf{z}}^{\cl}$. 
The equilibrium state of the quantum-classical system is thus described by the extended density matrix
\begin{equation}
\label{eq: extended rho}
    \rho_{\text{eq}}=\exp{\Big[-\beta\big(H_{x}^{\qu}+H_{\mathbf{z}}^{\cl}-\mathcal{F}_{\mathbf{z}}^{\cl}\big)\Big]}
     + \Big(\rho_{x}^{(x)}p+\rho_{x}^{(2x)}p^{2}-\rho_{x}^{(p)}\big(U_{x}'+\mathcal{F}_{x}'\big)\Big)P_{\text{eq}},
\end{equation}
where the first term $\exp{[-\beta\big(H_{x}^{\qu}+H_{\mathbf{z}}^{\cl}-\mathcal{F}_{\mathbf{z}}^{\cl}\big)]} = \varrho_{x}^{\qu}P_{\text{eq}}$ describes the quasi-static limit and the corrections in the second one account for the finite relaxation time of $\Sigma_{\qu}$.

\section{\label{sec:Stoch fw} Thermodynamic framework}

We now show how the dynamical framework developed in the previous section can be furnished with a consistent thermodynamic structure, first on the ensemble level, and then on the level of individual trajectories of the classical subsystem. 
In the second part, we focus on applications of our theory to hybrid thermal machines.

\subsection{Ensemble thermodynamics}

\subsubsection{First law}

Since the quantum-classical system is autonomous, the first law takes the form
\begin{equation}
\label{eq: 1st law}
    \dot{E}_t = \dot{Q}_{t}^{\qu} + \dot{Q}_{t}^{\cl},
\end{equation}
where $E_t$ denotes the internal energy of the total system and $\dot{Q}^{\qu/\cl}_t$ corresponds to the rate of heat uptake from the reservoir $R_{\qu/\cl}$. 
To derive explicit expressions for these quantities, we identify the internal energy of the quantum-classical system as
\begin{equation}\label{eq: int energy}
    E_t=\langle H_{x}^{\qu} + H_{\mathbf{z}}^{\cl}\rangle_t.
\end{equation}
The total rate of heat uptake then becomes
\begin{equation}
\label{eq: 1st law h uptake}
    \dot{E}_t =\int d\mathbf{z} \ \Big\{\text{Tr}\big[H_{x}^{\qu}\dot{\rho}_{\mathbf{z},t}\big]+H_{\mathbf{z}}^{\cl}\dot{P}_t\Big\}
    \simeq\int d\mathbf{z} \ \Big\{\text{Tr}\big[H_{x}^{\qu}\psi_{\mathbf{z}}\big]+H_{\mathbf{z}}^{\cl}\Big\}\dot{P}_t,
\end{equation}
where we have introduced the abbreviation
\begin{equation}
    \psi_{\mathbf{z}}\equiv\varrho_{x}^{\qu} +\rho_{x}^{(x)}p
\end{equation}
and the time derivative of the extended density matrix has been taken in first order in the adiabatic expansion \eqref{eq: rho ad exp} so that $\dot{E}_t$ is consistently obtained to second order. Upon inserting the Fokker-Planck equation \eqref{eq: FP eq} into \eqref{eq: 1st law h uptake} and integrating by parts with respect to $x$ and $p$, we find
\begin{align}
\label{eq: final Edot}
    \dot{E}_t&=\int d\mathbf{z} \ \bigg\{\text{Tr}\big[H_{x}^{\qu}\Lin{x}\rho_{\mathbf{z},t}\big]+\big(\zeta_{x}^{\qu}p+ U_{x}'+\mathcal{F}_{x}'\big)j^x_t +\frac{p}{m}j^p_t \bigg\}
    \\
    &=\int d\mathbf{z} \ \bigg\{\text{Tr}\big[H_{x}^{\qu}\Lin{x}\rho_{\mathbf{z},t}\big]-\frac{\zeta^{\cl}p^2+D^{\cl}+D_{x}^{\qu}}{m}P_t\bigg\},\nonumber
\end{align}
where we have used the relation
\begin{equation}
    \label{eq:34}j^x_t\partial_x\psi_{\mathbf{z}}+j^p_t\partial_p\psi_{\mathbf{z}}=\Lin{x}\rho_{\mathbf{z},t}.
\end{equation}
This identity can be derived by first recalling the Eqs.~\eqref{eq: 2nd order rho} and \eqref{eq: rho ad exp}, which imply 
\begin{equation}
\mathsf{L}_x \rho_{\mathbf{z},t} = (\varrho^{\qu'}_x + p \rho^{(x)\prime}_x)j^x_t + \varrho^{(x)}_x j^p_t
\end{equation}
with $j^x_t$ and $j^p_t$ defined in Eqs.~\eqref{eq: prob currents}, and, second, noting that $\partial_x \psi_\mathbf{z} = \varrho^{\qu\prime}_x + p \rho^{(x)\prime}_x$ and $\partial_p \psi_{\mathbf{z}} = \rho^{(x)}_x$.
The final expression in Eq.~\eqref{eq: final Edot} follows by inserting Eq.~\eqref{eq: prob currents} and integrating by parts with respect to $p$. 
This result leads us to identify $\dot{Q}_{t}^{\qu}$ and $\dot{Q}_{t}^{\cl}$ as
\begin{subequations}
\label{eq: q currents}
    \begin{align}
        \dot{Q}_{t}^{\qu}&=\int d\mathbf{z} \ \bigg\{\text{Tr}\big[H_{x}^{\qu}\Lin{x}\rho_{\mathbf{z},t}\big]+\frac{D_{x}^{\qu}}{m}P_t\bigg\},\label{subeq: q currents a}
        \\
        \dot{Q}_{t}^{\cl}&=\int d\mathbf{z} \ \Big\{D^{\cl} - \zeta^{\cl}p^2\Big\}\frac{P_t}{m}.
    \end{align}
\end{subequations}
It is straightforward to check, by substituting Eq.~\eqref{eq: extended rho} into these identifications, that both the quantum and classical heat currents vanish in equilibrium, i.e., for $\betatx{qu}=\betatx{cl}$.  

\subsubsection{Second law}\label{sec: ens 2nd law}

To show that the expressions \eqref{eq: q currents} for the heat currents are consistent with the second law, we identify the entropy of the hybrid system as \cite{Alonso2020}
\begin{equation}
    S_t=-\int d\mathbf{z} \ \text{Tr}\big[\rho_{\mathbf{z},t}\ln{\rho_{\mathbf{z},t}}\big].
\end{equation}
The rate of entropy production in the system then becomes
\begin{align}
\label{eq: S rate1}
    \dot{S}_t=-\int d\mathbf{z} \ \text{Tr}\bigl[\dot{\rho}_{\mathbf{z},t}\ln \rho_{\mathbf{z},t}\big]
    	& \simeq - \int d\mathbf{z} \ \text{Tr}
    		\bigl[(\varrho^{\qu}_x + p \rho^{(x)}_x)\ln \rho_{\mathbf{z},t} \bigr]\dot{P}_t\\
    	& = \int d\mathbf{z} \ \text{Tr}\bigl[\psi_\mathbf{z}\ln\rho_{\mathbf{z},t} \bigr]
    		\bigl(\partial_x j^x_t + \partial_p j^p_t\bigr)\nonumber\\
   		& = - \int d\mathbf{z} \ (j^x_t\partial_x + j^p_t \partial_p)
   			\text{Tr}\bigl[\psi_\mathbf{z}\ln\rho_{\mathbf{z},t} \bigr]\nonumber\\
    	& =-\int d\mathbf{z} \ \text{Tr}\big[\big(\Lin{x}\rho_{\mathbf{z},t}\big)
    		\ln{\rho_{\mathbf{z},t}}+\psi_{\mathbf{z}}\big(j_{t}^{x}\partial_x + j_{t}^{p}\partial_{p}\big)		
    			\ln{\rho_{\mathbf{z},t}}\big],\nonumber
\end{align}
Here, we have inserted the first-order truncation of the adiabatic-response expansion \eqref{eq: rho ad exp} of $\rho_{\mathbf{z},t}$ and the Fokker-Planck equation \eqref{eq: FP eq}; 
the third line of Eq.~\eqref{eq: S rate1} then follows from an integration by parts with respect to the phase space variables and the final result is obtained by applying the product rule for the partial derivatives and inserting the identity \eqref{eq:34}.
Upon expanding to second order in the adiabatic perturbation series, we thus obtain
\begin{equation}
\label{eq: Sdot}
    \dot{S}_t\simeq\betatx{qu}\dot{Q}_{t}^{\qu}+\int d\mathbf{z} \ \bigg\{\frac{\betatx{qu}}{m}\big(\zeta_{x}^{\qu}p^2-D_{x}^{\qu}\big)P_t
    -\frac{\big(\mathcal{j}_{t}^{\qu}+\mathcal{j}_{t}^{\cl})\partial_{p}P_t}{P_t}\bigg\}.
\end{equation}
Here we have introduced the irreversible probability currents
\begin{subequations}
\label{eq: irr prob cur}
\begin{align}
    \mathcal{j}_{t}^{\qu}&\equiv-\zeta_{x}^{\qu}pP_t-D_{x}^{\qu}\partial_{p}P_t,
    \\  
    \mathcal{j}_{t}^{\cl}&\equiv-\zeta^{\cl}pP_t-D^{\cl}\partial_{p}P_t,
    \end{align}
\end{subequations}
and inserted the expression \eqref{subeq: q currents a} for the quantum heat current $\dot{Q}_{t}^{\qu}$. Furthermore, we have used the relation
\begin{equation}
    \partial_{\alpha}\ln{Y_\alpha}=\mathsf{T}\big[Y_\alpha\big]\partial_\alpha Y_\alpha,
\end{equation}
with the superoperator
\begin{equation}
    \mathsf{T}\big[Y\big]\circ=\int_0^\infty d\lambda \ \big(Y+\lambda\big)^{-1}\circ\big(Y+\lambda\big)^{-1},
\end{equation}
which holds for any positive definite matrix $Y$ \cite{Lieb1973}. Finally, inserting the definitions \eqref{eq: irr prob cur} of the dissipative currents into Eq.~\eqref{eq: Sdot} gives
\begin{equation}
    \dot{S}_t=\betatx{qu}\dot{Q}_{t}^{\cl}+\betatx{cl}\dot{Q}_{t}^{\cl}+\sigma_t
\end{equation}
with the total rate of entropy production
\begin{equation}\label{eq: rate ent prod}
    \sigma_t\equiv \int d\mathbf{z} \ \Bigg\{\frac{\big(\mathcal{j}_{t}^{\qu}\big)^2}{D_{x}^{\qu}P_t}+\frac{\big(\mathcal{j}_{t}^{\cl}\big)^2}{D^{\cl}P_t}\Bigg\}\geq0.
\end{equation}
This result shows that the heat currents defined in Eq.~\eqref{eq: q currents} are consistent with the second law,
\begin{equation}
\label{eq: 2nd law}
    \sigma_t=\dot{S}_t-\betatx{qu}\dot{Q}_{t}^{\qu}-\betatx{cl}\dot{Q}_{t}^{\cl}\geq0.
\end{equation}
We note that the irreversible currents $\mathcal{j}_{t}^{\qu/\cl}$ are zero for $\betatx{cl}=\betatx{qu}$ and hence the total rate of entropy production vanishes in equilibrium along with both heat currents, as thermodynamic consistency requires.

\newcommand{\qd}{\dot{q}_{t}^{\text{d}}}
\newcommand{\qex}{\dot{q}_{t}^{\text{ex}}}

\subsection{Stochastic thermodynamics}

On the ensemble level, the first and second law for quantum-classical hybrid systems are given by the Eqs.~\eqref{eq: final Edot} and \eqref{eq: 2nd law}. 
With these prerequisites, we are now ready to develop our stochastic thermodynamics at the quantum-classical boundary. 
To this end, we assume that the motion of the classical degree of freedom can be continuously monitored and thus provides a natural and physically transparent notion of single trajectories.

\subsubsection{Stochastic first law}

A stochastic first law can be derived from the expression
\begin{equation}
    \mathcal{E}^{\qu}[\mathbf{z}_t,\dot{\mathbf{z}}_t]\equiv \mathcal{E}_t^{\qu} = \text{Tr}\big[H_{x_t}^{\qu}\rho_{t}\big]
    \label{eq: stoch 1st law}
\end{equation}
for the trajectory-resolved internal energy of the quantum degrees of freedom. 
We recall that the density matrix of the quantum subsystem $\rho_t$ is given by Eq.~\eqref{eq: rho ad resp} and is thus a function of the classical phase-space variables and their time derivatives. 
By taking a time derivative of Eq.~\eqref{eq: stoch 1st law} and neglecting third-order corrections in the adiabatic-response expansion, we obtain the balance equation
\begin{align}
\label{eq: stoch 1st}
    \dot{\mathcal{E}}_{t}^{\qu} =-\text{Tr}\big[F_{x_t}\rho_t\big]p_t/m+\text{Tr}\big[H_{x_t}^{\qu}\dot{\rho}_t\big]
    & \simeq \mathcal{F}_{x_t}'p_t/m + \zeta_{t}^{\qu}p_t^2/m + \text{Tr}\big[H_{x_t}^{\qu}\Lin{x_t}\rho_t\big]\\
    & \equiv \dot{w}_t+\qex+\qd
    \nonumber
\end{align}
with
\begin{equation}
    \dot{w}_t=\mathcal{F}_{x_t}'p_t/m, \qquad
        \qex=\zeta_{t}^{\qu}p_t^2/m, \qquad \label{eq: dot qex}
        \qd=\text{Tr}\big[H_{x_t}^{\qu}\Lin{x_t}\rho_t\big].
\end{equation}
Here, we have identified $\dot{w}_t$ as the rate at which work is performed on the quantum system by the classical one along a given trajectory $\mathbf{z}_t$ and $\dot{q}^{\text{d}}_t$ denotes the associated heat flux from the quantum reservoir into the working medium. 
The quantity $\dot{q}^{\text{ex}}_t$ must then correspond to the rate at which heat is dissipated from the classical subsystem into the quantum one.

By taking the ensemble average, which we denote by $\langle\cdots\rangle_\text{en}$, the identification \eqref{eq: stoch 1st law} of the stochastic internal energy of the quantum subsystem leads to a natural division of the mean internal energy of the whole system into a quantum and a classical contribution, $E^{\qu}_t$ and $E^{\cl}_t$, which are given by 
\begin{equation}
\langle\mathcal{E}^{\qu}_t\rangle_\text{en} = \langle H^{\qu}_x\rangle_t\equiv E^{\qu}_t = E_t -  E^{\cl}_t \quad\text{and}\quad E^{\cl}_t \equiv \langle H^{\cl}_{\mathbf{z}}\rangle_t. 
\end{equation}
Here, we use average $\langle\cdots\rangle_t$ with respect to the extended density matrix, which defined in Eq.~\eqref{eq: full av}.
Taking the ensemble average of the stochastic first law \eqref{eq: stoch 1st} yields 
\begin{equation}\label{eq: av stoch 1st}
	\langle\dot{\mathcal{E}}^{\qu}_t\rangle_\text{en}
	= \dot{W}_t + \dot{Q}^\text{ex}_t + \dot{Q}^\text{d}_t,
\end{equation}
where 
\begin{equation}
\dot{W}_t \equiv \langle\dot{w}_t\rangle_\text{en}= \langle \mathcal{F}'_x p /m \rangle_t, \;\;\;
	\dot{Q}^\text{ex}_t \equiv \langle \dot{q}^\text{ex}_t \rangle_\text{en}
		=\langle \zeta^{\qu}_x p^2/m\rangle_t, \;\;\;
	\dot{Q}^\text{d}_t \equiv \langle \dot{q}^\text{d}_t\rangle_\text{en}
		=\langle \mathsf{L}^\ddagger_x H^{\qu}_x\rangle_t.
\end{equation}
Upon inserting the Eqs.~\eqref{eq: q currents} and noting that the Fokker-Planck equation \eqref{eq: FP eq} implies $\dot{E}^\text{cl}_t = -\langle \mathcal{F}'_x p/m +(\zeta^{\cl} + \zeta^{\qu}_x)p^2/m - (D^{\cl} + D^{\qu}_x)/m\rangle_t$, it can be shown explicitly that Eq.~\eqref{eq: av stoch 1st} is equivalent to the identity
\begin{equation}
	\dot{E}^{\qu}_t =\dot{Q}^{\cl}_t +\dot{Q}^{\qu}_t -\dot{E}^{\cl}_t.
\end{equation}
Hence, the stochastic first law \eqref{eq: stoch 1st law} is indeed consistent with its ensemble version \eqref{eq: 1st law}.

\subsubsection{Stochastic second law}

A plausible identification of the stochastic entropy of the quantum subsystem is given by the von Neumann entropy of its trajectory-resolved density matrix,
\begin{equation}\label{eq: stoch ent}
    \mathcal{S}^{\qu}[\mathbf{z}_t,\dot{\mathbf{z}}_t]\equiv\mathcal{S}_{t}^{\qu}=-\text{Tr}\big[\rho_t\ln{\rho_t}\big].
\end{equation}
Taking a time derivative of this quantity and neglecting third-order corrections gives
\begin{equation}
\label{eq: stoch 2nd}
    \dot{\mathcal{S}}^{\qu}_t\simeq-\text{Tr}\big[(\Lin{x_t}\rho_{t} )\ln{\rho_t}\big]
    =\text{Tr}\big[(\Lin{x_t}\rho_t )(\ln{\varrho_{x_t}^{\text{qu}}}-\ln{\rho_t})\big]+\betatx{qu}\dot{q}^{\text{d}}_t,
\end{equation}
where $\varrho^{\text{qu}}_{x_t}$ is the instantaneous Gibbs state of the quantum subsystem defined in Eq.~\eqref{eq: Gibss state}. 
Since $\Lin{x}\varrho^{\text{qu}}_x=0$, the first term in the second line is non-negative by Spohn's theorem \cite{Spohn1978}. 
For fixed $\betatx{qu}$, we therefore have the stochastic second law
\begin{equation}\label{eq: stoch 2nd law}
    \dot{\mathcal{S}}^{\qu}_t - \beta^{\qu}\dot{q}^{\text{d}}_t\geq 0.
\end{equation}

As in the case of the internal energy, taking the ensemble average of the stochastic entropy \eqref{eq: stoch ent} induces a natural division of the total entropy of the combined system into a quantum and a classical contribution, which we call $S^{\qu}_t$ and $S^{\cl}_t$. 
Up to second-order corrections in the adiabatic expansion, these quantities are given by 
\begin{equation}
	\langle \mathcal{S}^{\qu}_t\rangle_\text{en} \simeq -\langle \ln \psi_{\mathbf{z}}\rangle_t 
		\equiv S^{\qu}_t \simeq S_t - S^{\cl}_t 
		\quad\text{and}\quad
		S^{\cl}_t \equiv - \int d\mathbf{z} \; P_t \ln P_t. 
\end{equation}
Averaging the stochastic second law \eqref{eq: stoch 2nd law} yields the inequality 
\begin{equation}\label{eq: qu 2nd law}
	\langle \dot{\mathcal{S}}^{\qu}_t \rangle_\text{en} - \beta^{\qu} \dot{Q}^\text{d}_t \geq 0,
\end{equation}
which is stronger the ensemble-level second law \eqref{eq: 2nd law} in that it involves only the quantum part of the system entropy\footnote{
The inequality \eqref{eq: qu 2nd law} can be explicitly verified on the ensemble level; by following the steps of Sec.~\ref{sec: ens 2nd law}, one finds $\langle\dot{\mathcal{S}}^{\qu}_t\rangle_\text{en} -\beta^{\qu}\dot{Q}^\text{d}_t \simeq \beta^{\qu}\langle \zeta^{\qu}_x p^2/m\rangle \geq 0$, where third-order corrections in the adiabatic expansion have to be neglected.
However, the physical meaning of the quantities $\langle \dot{S}^{\qu}_t\rangle_\text{en}$ and $\dot{Q}^\text{d}_t$ becomes clear only on the trajectory level.}. 
Finally, by following the derivation of Sec.~\ref{sec: ens 2nd law}, one can explicitly show that 
\begin{equation}
	\dot{S}^{\qu}_t \simeq \beta^{\qu}\dot{Q}^{\qu}_t + \beta^{\cl} \dot{Q}^{\cl}_t + \sigma_t - \dot{S}^{\cl}_t
\end{equation}
in second order of the adiabatic expansion, where the total rate of entropy production $\sigma_t \geq 0$ was defined in Eq.~\eqref{eq: rate ent prod}. 
Thus, we have recovered the ensemble-level second law \eqref{eq: 2nd law} from its stochastic counterpart \eqref{eq: stoch 2nd law}. 

\subsubsection{Engine cycles}

We are now ready to formulate the stochastic thermodynamics of quantum thermal machines, which deliver output to a classical load. 
To this end, we consider the quantum and classical subsystems, respectively, as the working medium and load of such a machine. 
Input is provided in the form of thermal energy from a heat source, which periodically modulates the temperature of the quantum reservoir $R_{\qu}$, thus driving the joint system into a non-equilibrium periodic state.
For the sake of concreteness we focus on a two-stroke engine cycle, whereby the inverse temperature of the quantum reservoir switches periodically between two values $\beta^{\qu}_{h}$ and $\betatx{qu}_c>\betatx{qu}_h$. 
The switching takes place at the time $0<t_s<\tau$ where $\tau$ denotes the period of the cycle. 
Thus, up to a short time interval immediately after each temperature switch, where the quantum subsystem is no longer close to its instantaneous equilibrium state, the cycle consists of two irreversible isothermal processes, or strokes, each of which can be described within our adiabatic response approach.
Furthermore, any corrections to the dissipated heat that arise from the relaxation processes between the strokes can be consistently evaluated as a difference between instantaneous equilibrium energies, as long as the temperature gradient is sufficiently small\footnote{More specifically, we assume that $\mathcal{E}^{\Sigma_\qu}(\beta_c^\qu-\beta_h^\qu)\sim \tau^{\Sigma_{\cl}}/\tau^{\Sigma_\qu}\equiv \varphi$, where $\mathcal{E}^{\Sigma_\qu}$ is the characteristic energy scale of the quantum subsystem and $\tau^{\Sigma_{\cl}}$ and $\tau^{\Sigma_{\qu}}$ are the characteristic time scales of the classical and quantum degrees of freedom; hence $\varphi$ can be regarded as the small parameter of the adiabatic expansion and we calculate the accumulated stochastic quantities \eqref{eq: ext stoch quant} and \eqref{eq: qh,qc} to first order in this parameter. 
Since the switching heat \eqref{eq: vareps} if of first order in $\mathcal{E}^{\Sigma_\qu}(\beta^\qu_c - \beta^\qu_h)\sim\varphi$, it can thus be evaluated consistently with respect to the instantaneous equilibrium states of the quantum subsystem, that is, in zeroth order of the adiabatic expansion, for details see Refs.~\cite{Eglinton2022} and \cite{Menczel2019}.}.
That is, the accumulated stochastic work and heat contributions for one cycle are given by 
\begin{equation}\label{eq: ext stoch quant}
    w=\int_0^\tau dt \ \dot{w}_t,
    \qquad
    q^{\text{ex}}=\int_0^{\tau}dt \ \dot{q}^{\text{ex}}_t,
    \qquad
    q^{\text{d}}=q^{\text{d}}_h+q^{\text{d}}_c,
\end{equation}
where
\begin{equation}
    q^{\text{d}}_h=\int_0^{t_s}dt \ \dot{q}^{\text{d}}_t +\varepsilon_{x,0},
    \qquad q^{\text{d}}_c=\int_{t_s}^{\tau}dt \ \dot{q}^{\text{d}}_t + \varepsilon_{x,t_s}
    \label{eq: qh,qc}
\end{equation}
and 
\begin{equation}
\label{eq: vareps}
    \varepsilon_{x,t}= \lim_{\epsilon\to 0} \Big(\text{Tr}\big[H^{\qu}_{x,t+\epsilon}\varrho_{x,t+\epsilon}^{\qu}\big]-\text{Tr}\big[H^{\qu}_{x,t-\epsilon}\varrho_{x,t-\epsilon}^{\qu}\big]\Big)
\end{equation}
account for the heat that is dissipated while the quantum subsystem returns to its adiabatic-response state right after the temperature switches.

Integrating the first and second law, Eqs.~\eqref{eq: stoch 1st} and \eqref{eq: stoch 2nd}, over a full period therefore provides the relations
\begin{subequations}
    \begin{align}
        \mathcal{E}_{\tau}^{\qu} - \mathcal{E}_{0}^{\qu} &= w  + q^{\text{d}}  +q^{\text{ex}},
        \\
        \mathcal{S}_{\tau}^{\qu} - \mathcal{S}_{0}^{\qu} &\geq\betatx{qu}_{h}q^{\text{d}}_h + \betatx{qu}_{c}q^{\text{d}}_c,
    \end{align}
\end{subequations}
where the differences on the left do not vanish in general, even though the system is in a periodic state, since $\mathcal{E}^{\qu}$ and $\mathcal{S}^{\qu}$ are stochastic variables. 
After taking the ensemble average, however, we are left with
\begin{subequations}
\label{eq: stoch ensmb 1st and 2nd}
\begin{align}
\label{eq: ensemb 1st}
        0&=W+Q^{\text{ex}}_h+Q^{\text{ex}}_c+Q^{\text{d}}_h+Q^{\text{d}}_c,
        \\
       0&\geq \betatx{qu}_{h}Q_{\text{h}}^{\text{d}} + \betatx{qu}_{c}Q_{\text{c}}^{\text{d}}.\label{eq: ensemb 2nd}
\end{align}
\end{subequations}
Here, 
\begin{equation}
\label{eq: stoch ensmb W}
    W= \int_{0}^{\tau}dt \ \dot{W}_t
\end{equation}
corresponds to the average work that is delivered to the classical degree of freedom per cycle and
\begin{subequations}
\label{eq: stoch ensmb Q}
    \begin{align}
    Q^{\text{ex}}_h &= \int_{0}^{t_s}dt \ \dot{Q}^\text{ex}_t,  
    \label{eq: stoch ensmb Q a}
        \\
        Q^{\text{d}}_h &=\int_{0}^{t_s}dt \  \dot{Q}^\text{d}_t
        +\int d\mathbf{z} \ \varepsilon_{x,0},
    \end{align}
\end{subequations}
denote the mean heat exchange between working system and load and the mean heat uptake from the quantum reservoir during the hot phase of the cycle, respectively; the corresponding quantities $Q^{\text{ex}}_c$ and $Q^{\text{d}}_c$ for the cold stroke are defined analogously. 
The efficiency of the engine can now be consistently defined as
\begin{equation}
    \eta=-W/Q_{\text{h}}^{\text{d}}\leq1+Q_{\text{c}}^{\text{d}}/Q_{\text{h}}^{\text{d}}\leq\eta_{\text{C}}
\end{equation}
where the upper bound $\eta_{\text{C}}=\betatx{qu}_h/\betatx{qu}_c$, which corresponds to the Carnot value, follows directly from the relations \eqref{eq: stoch ensmb 1st and 2nd} and by noting that $Q^{\text{ex}}_h,Q^{\text{ex}}_c\geq0$, as becomes evident from Eq.~\eqref{eq: stoch ensmb Q a}, since $\zeta^{\qu}_x$ and $P_t$ are both non-negative.

\section{\label{sec:Rydb eng} Quantum-classical Rydberg system}

After developing our general framework, we will now show how it can be applied to a concrete model of a heat engine based on an opto-mechanical setup. 
We first outline the physical setup and then develop its mathematical description with the help of suitable approximations. 
Specifically, we will consider two examples of thermodynamic processes, a single-stroke expansion and a two-stroke cycle. 
We then analyze the statistics of thermodynamic quantities such as work and heat and calculate the efficiency of the two-stroke engine by means of numerical simulations.

\newcommand{\omegaa}{\omega_{\text{a}}}
\newcommand{\Hqu}{H_{x}^{\text{qu}}}
\newcommand{\sumN}[2]{\sum_{#1=1}^{#2}}

\subsection{System}

\begin{figure}[t]
    \centering
    \includegraphics[width=.6\textwidth]{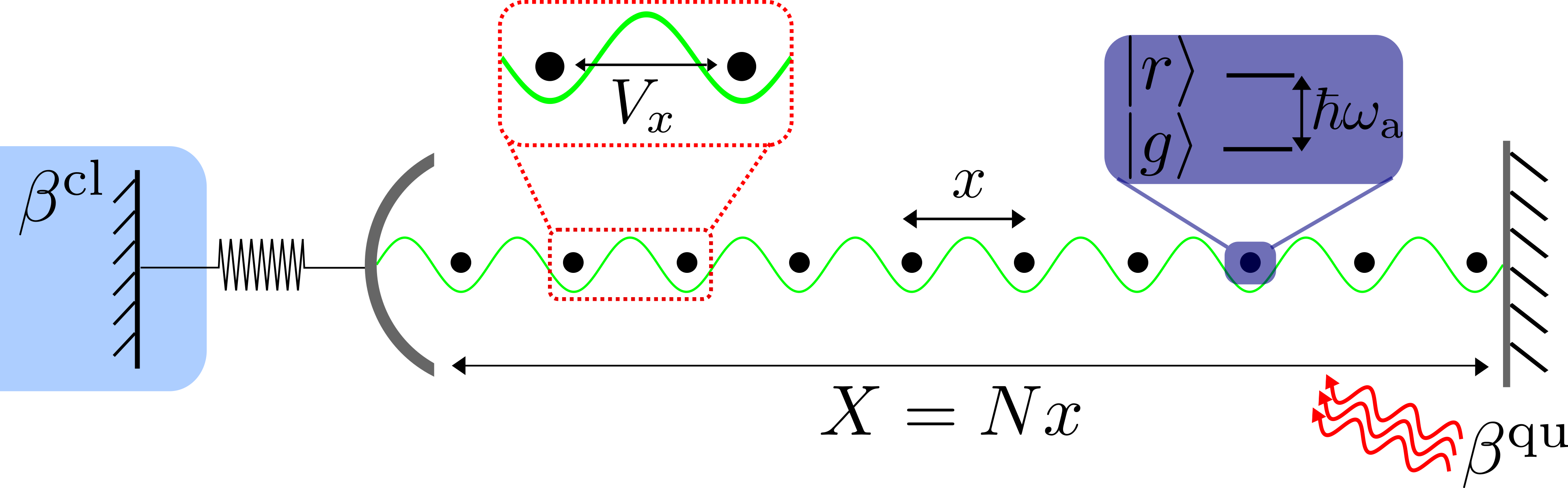}
    \caption{Schematic of our quantum-classical engine. 
    	The working system consists of a chain of $N$ Rydberg atoms, which sit at the minima 
    	of a standing light wave inside an optical cavity. 
    	The atoms are modelled as two-level systems and are equally spaced by a distance $x$, 
    	which effectively plays the role of the classical degree of freedom. 
    	The cavity length is $X=Nx$. 
    	The atoms interact via a dipole-dipole interaction, described by the potential $V_x$ 
    	and are driven by a noisy laser that mimics a thermal reservoir with inverse 
    	temperature $\betatx{qu}$. 
    	The classical work load consists of a movable mirror, which is coupled to a classical 
    	bath with inverse temperature $\betatx{cl}$. 
    	The motion of the mirror is driven by the radiation pressure inside the cavity, 
    	which depends on the state of the atoms and the restoring force of a classical spring 
    	with constant $c$.}
    \label{fig: Rydb quantum-classical eng}
\end{figure}

We consider the hybrid engine shown in Fig.~\ref{fig: Rydb quantum-classical eng}. 
A chain of $N$ interacting Rydberg atoms are confined in an optical cavity by a lattice potential created by a standing light wave. 
Each Rydberg atom is treated as a two-level system (TLS), with ground state $|g\rangle$ and excited Rydberg state $|r\rangle$. 
We assume that the two levels are coupled by a noisy laser, so that quantum superpositions between the two states decohere quickly. State changes within the atoms can thus be described by a rate equation. 
The noisy laser plays the role of a thermal reservoir\footnote{
Alternatively, one could imagine the thermal reservoir to be mimicked by a second cavity, which is arranged orthogonal to the engine cavity, tuned in resonance with the Rydberg transition and driven by an additional laser. 
Noise and dephasing could then be introduced by stochastically changing the position of one of the mirrors of this second cavity. 
As a result, the excitation and de-excitation processes of the Rydberg atoms induced by the second cavity would become quasi-classical.
}, and can be modelled as an ensemble of non-interacting bosons at inverse temperature $\betatx{qu}$ \cite{Marcuzzi2016}. 
The free evolution of the Rydberg chain is described by the Hamiltonian
\begin{equation}
    \label{eq: Hqu}
    \Hqu = \hbar\omegaa \sumN{k}{N} n_k + V_{x}\sumN{k}{N-1}  n_k n_{k+1},
\end{equation}
where $n_{k}=|r_k\rangle\langle r_k|$ and the atoms experience nearest neighbour dipole-dipole interactions $V_{x}=v/x^{3}$, which depend on their separation $x$. 
The cavity is formed by a fixed mirror and a movable mirror of mass $m$, which is in contact with a classical reservoir with inverse temperature $\betatx{cl}$. 
The position of the movable mirror plays the role of the classical degree of freedom, $X_t = Nx_t$, and determines the size of the cavity and thus the length of the standing light wave that confines the atomic array. 
That is, we assume that the atoms sit at the minima of the trapping potential and are therefore evenly spaced along the cavity.
The quantum and classical subsystems interact with each other through the radiation pressure inside the cavity, which effectively arises from the interactions between neighbouring atoms. 
The resulting force on the movable mirror is counterbalanced by the restoring force of a classical spring
\begin{equation}
    -\partial_{X}U_{X}=-c\big(X-X_0\big),
\end{equation}
with spring constant $c$, which pushes the mirror to its natural position $X_0$.

\subsection{Model}

\subsubsection{Quantum subsystem}

The state of the quantum system $\rho_t$ evolves under a collective many-body master equation,
\begin{equation}
\label{eq: ME ryd}
    \dot{\rho}_t = \Lin{x}\rho_t\equiv\Hin{x}\rho_t+\Din{x}\rho_t,
\end{equation}
where
\begin{equation}
\Hin{x}\circ = -\frac{i}{\hbar}\big[\circ,\Hqu\big]\quad\text{and}\quad
\Din{x}\circ=\sumN{k}{N}\sum_{\alpha=0}^{2}\Big\{\chi_{x}^{\alpha-}\Kin{J_{k}^{\alpha}}\circ+\chi_{x}^{\alpha+}\Kin{J_{k}^{\alpha\dagger}}\circ\Big\}
\end{equation}
denote the Liouville and dissipative super-operators and $\Kin{Y}$ is defined in Eq.~\eqref{eq: kin}. 
The energy of emitted and absorbed photons from a particular atom depend on the state of the neighbouring atoms in the chain, owing to the nearest neighbour interaction \cite{Nill2022}. 
The collective jump operators that describe this effect are given by
\begin{equation}
    J^{\alpha}_{k}=\begin{cases}
        \tilde{n}_{k-1}\sigma_k^-\tilde{n}_{k+1} & \text{if $\alpha=2$} \\
        (\mathbb{I}_k-\tilde{n}_{k-1})\sigma_k^-\tilde{n}_{k+1}
        + \ \tilde{n}_{k-1}\sigma_k^-(\mathbb{I}_k-\tilde{n}_{k+1})  & \text{if $\alpha=1$} \\
        (\mathbb{I}_k - \tilde{n}_{k-1})\sigma_k^-(\mathbb{I}_k - \tilde{n}_{k+1}) & 
        \text{if $\alpha=0$}
    \end{cases},
\end{equation}
with $\tilde{n}_{k-1}\equiv n_{k-1}(1-\delta_{k-1})$ and $\tilde{n}_{k+1}\equiv n_{k+1}(1-\delta_{k-N})$. 
That is, the jump operator $J^\alpha_k$ induces the decay of the $k^\text{th}$ atom in the chain from its Rydberg state to its ground state, if $\alpha$ of its neighbors are excited;
the adjoint jump operator $J^{\alpha\dagger}_k$ describes the reverse process, i.e., the excitation of the $k^\text{th}$ atom in the presence of $\alpha$ excited neighbors.
Since we consider open boundary conditions, only the jump operators with $\alpha\leq1$ are defined at the ends of the chain, i.e., for $k=1$ and $k=N$. Furthermore,
\begin{equation}
    \mathbb{I}_k=n_{k}+g_{k} \ \ \text{with} \ \ g_k=|g_k\rangle\langle g_k|
\end{equation}
denotes the single-atom identity matrix. 
The resulting Bohr frequencies account for the number of excited atoms $\alpha$ in the neighbourhood of the $k$-th atom and read
\begin{equation}
    \nu_{x,\alpha} = \omegaa + \alpha V_{x}/\hbar.
\end{equation}
The corresponding collective decay rates are
\begin{equation}
    \chi_{x}^{\alpha+}=\gamma_{x,\alpha}\mathcal{B}_{x,\alpha} \ \ \text{and} \ \ \chi_{x}^{\alpha-}=\gamma_{x,\alpha}\big(\mathcal{B}_{x,\alpha}+1\big),
\end{equation}
where $\gamma_{x,\alpha}=2\pi\hbar\kappa\nu_{x,\alpha}^{3}$. 
The constant $\kappa$ sets the scale of the decay rates and the Bose-Einstein distribution
\begin{equation}
    \mathcal{B}_{x,\alpha}=\big(\exp[\betatx{qu}\hbar\nu_{x,\alpha}]-1\big)^{-1}
\end{equation}
describes the effective reservoir mimicked by the noisy laser. Thus, the collective decay rates obey the detailed balance condition
\begin{equation}
    \chi_{x}^{\alpha+}/\chi_{x}^{\alpha-}=\exp[-\betatx{qu}\hbar\nu_{x,\alpha}]
\end{equation}
and the unique stationary solution of the master equation \eqref{eq: ME ryd} is the Gibbs state
\begin{equation}
    \varrho_{x}^{\qu}=\exp\big[-\betatx{qu}(H_{x}^{\qu}-\mathcal{F}_{x})\big],
\end{equation}
where
\begin{equation}
    \mathcal{F}_{x}\equiv-\ln {\text{Tr}\big[\exp[-\betatx{qu}H_{x}^{\qu}]\big]}/\betatx{qu}
\end{equation}
is the free energy of the quantum system. The non-local master equation \eqref{eq: ME ryd}
has been derived following the approach taken in Ref.~\cite{Nill2022}, and adapting it to allow for a finite reservoir temperature.

\subsubsection{Classical subsystem}

The classical equations of motion are the Langevin equations
\begin{subequations}
\label{eq: classicRyd langevin}
\begin{align}
    \dot{X}_{t} &= Np_t/m,\\
    N\dot{p}_t &=  -\partial_{X_t} U_{X_t} + f_{\mathbf{z}_t,t}^{\qu} -N\zeta^{\cl}p_t + \xi_{t}^{\cl},
\end{align}
\end{subequations}
where $p_t$ is the momentum of a single atom of the chain and $f_{\mathbf{z}}^{\qu}$ is the quantum coupling force. 
Since the position of the movable mirror is linearly dependent on the inter-atomic spacing we can, upon rescaling with the atom number $N$, replace the cavity length $X$ with the spacing between atoms $x$ as our classical degree of freedom. 
The logic behind this rescaling is two-fold. 
First, since the atoms are evenly spaced inside the cavity, working with the interatomic spacing will allow us to apply the equations of motion for an arbitrary number of atoms, where we have an equilibrium spacing $x_0=X_0/N$ corresponding to an experimentally feasible value. 
That is, we increase the equilibrium length of the cavity proportional to the number of atoms so that an infinite density scenario, where the spacing between atoms inside the cavity becomes unfeasibly small, is avoided. 
Second, the Hamiltonian of the quantum system and the generator of the master equation that describes its time evolution depend only on the interatomic spacing and so rescaling allows us to consistently work with a single classical degree of freedom. 
The time evolution of the spacing is governed by the Langevin equations
\begin{subequations}
\label{eq: Ryd langevin}
\begin{align}
    \dot{x}_{t} &= p_t/m,\\
    \dot{p}_t &= -U_{t}' + f_{\mathbf{z}_t,t}^{\qu}/N -\zeta^{\cl}p_t + \xi_{t}^{\cl}/N,
\end{align}
\end{subequations}
where the restoring force of the movable mirror is now
\begin{equation}
    -U_{x}'=-c\big(x-x_{0}\big)
\end{equation}
with primes denoting partial derivatives with respect to $x$. 
The classical diffusion constant $D^{\cl}$ enters through the fluctuation-dissipation relation \eqref{eq: cl fluc dis} and the classical stochastic force $\xi_{t}^{\cl}$ obeys the relations \eqref{eq: cl xi}. 
The quantum coupling force per particle,
\begin{equation}
\label{eq: Ryd qf}
    f_{\mathbf{z},t}^{\qu}/N = -\mathcal{F}_{x}'/N^2 - \zeta_{x}^{\qu}p+\xi_{x,t}^{\qu}/N,
\end{equation}
contains the deterministic force per particle $-\mathcal{F}_{x}'/N^2$ and the friction force $-\zeta_{x}^{\qu}p$, where $\zeta_{x}^{\qu}$ satisfies the fluctuation-dissipation relation \eqref{eq: Dq fdr} and the associated diffusion constant is given by 
\begin{equation}
\label{eq: Ryd Dqu}
    D_{x}^{\qu}=-\llangle \big(\Lin{x}^{\ddagger}\big)^{-1}\delta F_{x} ; \delta F_{x}\rrangle_{\varrho_{x}^{\qu}}
\end{equation}
with
\begin{equation}
    \delta F_{x}=-\frac{V_x'}{N} \bigg(\sumN{k}{N}n_k n_{k+1} - \text{Tr}\big[\sumN{k}{N}n_k n_{k+1} \varrho_{x}^{\qu}\big]\bigg)
\end{equation}
for the Rydberg chain.
The stochastic force per particle $\xi^{\qu}_{x,t}/N$ describes the noise induced by the quantum subsystem and is related to the diffusion constant $D_x^{\qu}$ by Eqs.~\eqref{eq: qu fluc rel}. 

\newcommand{\nbar}{\langle n \rangle}
\newcommand{\mf}{\text{MF}}
\newcommand{\chimf}{\bar{\chi}}
\newcommand{\Xmf}{\bar{\chi}}
\newcommand{\Linmf}[1]{\bar{\mathsf{L}}_{#1}}
\newcommand{\Hinmf}[1]{\bar{\mathsf{H}}_{#1}}
\newcommand{\Dinmf}[1]{\bar{\mathsf{D}}_{#1}}
\newcommand{\brho}{\bar{\rho}}

\subsection{Mean-field approximation}

\subsubsection{Quantum subsystem}

Obtaining analytic expressions for the quantum coupling force is in general not straightforward for a many-body system. 
In particular, owing to the collective nature of the master equation \eqref{eq: ME ryd}, where jump rates and operators for individual atoms depend on their immediate neighbourhood, the diffusion constant \eqref{eq: Ryd Dqu} would have to be determined numerically, which is possible only for small values of $N$. 
To calculate mean-values of few body observables, however, we do not necessarily need to resolve the neighbourhood of each individual atom. 
Instead, we may assume that the interaction between neighbouring atoms induces, on average, a homogeneous energy shift. 
This assumption is justified if the scale of thermal fluctuations is large relative to the scale of the inter-atomic interactions, i.e., if $\betatx{qu}V_{x}\ll 1$ and if the number of atoms is sufficiently large so that individual local excitations have only a negligible effect on average quantities. 
If these conditions are met, we can proceed with a \textbf{mean-field (MF)} approach, where the nearest neighbour excitation operators $n_{k\pm1}$ in the collective master equation \eqref{eq: ME ryd} are replaced with the average excitation density
\begin{equation}
    \langle n_x\rangle=\frac{1}{N}\text{Tr}\biggl[\sumN{k}{N}n_k\varrho_{x}^{\qu}\biggr]\equiv\frac{1}{N}\partial_{\hbar\omegaa}\mathcal{F}_{x}.
\end{equation}
The MF master equation thus becomes
\begin{equation}
\label{eq: mf me}
    \dot{\rho}_t=\Linmf{x}\rho_{t}\equiv\Hin{x}\rho_t+\Dinmf{x}\rho_t,
\end{equation}
where
\begin{equation}
\label{eq: mf dissipator}
    \Dinmf{x}\circ=\sumN{k}{N}\Big\{\bar{\chi}_{x,k}^{-}\Kin{\sigma_{k}^{-}}\circ+\bar{\chi}_{x,k}^{+}\Kin{\sigma_{k}^{+}}\circ\Big\},
\end{equation}
and we use bars to denote MF quantities throughout. 
The MF rates reflect the collective nature of decay and excitation and are given by
\begin{equation}
\bar{\chi}^{\pm}_{x,k}=\begin{cases}
\nbar^{2}\chi^{1\pm}+(1-\nbar)^{2}\chi^{0\pm}& \text{if $k=1$ or $N$}\\
    \nbar^{4}\chi^{2\pm} + 4\nbar^{2}\big(1-\nbar\big)^{2}\chi^{1\pm}
    +(1-\nbar)^{4}\chi^{0\pm} & \text{otherwise}
\end{cases},
\end{equation}
We note that, for notational convenience, we drop the $x$-dependence of the collective rates $\bar{\chi}^{\alpha\pm}$ and the excitation density $\langle n \rangle$ from here onwards. 
The unique stationary solution of the master equation \eqref{eq: mf me} is given by the product state
\begin{equation}
\label{eq: mf ss}
    \bar{\varrho}_x^{\qu} = \bigotimes_{k=1}^{N}(\bar{\chi}_{x,k}^{+} n_k+\bar{\chi}_{x,k}^{-}g_k)/\Xmf_{x,k}^N\quad \text{with} \quad \Xmf=\bar{\chi}^{-}+\bar{\chi}^{+},
\end{equation}
which approximates the thermal state of the quantum subsystem in the parameter regimes, where the MF approximation is valid.

In addition we now replace open with periodic boundary conditions. 
Although our physical setup in principle requires open boundary conditions, this approximation, which enhances the transparency of our mathematical model, is still justified for sufficiently large chains, where boundary effects play only a minor role. 
The MF decay rates then become independent of $k$ and are given by 
\begin{equation}
    \chimf_{x}^{\pm}=\nbar^{4}\chi^{2\pm} + 4\nbar^{2}\big(1-\nbar\big)^{2}\chi^{1\pm}+(1-\nbar)^{4}\chi^{0\pm}.
\end{equation}
With these simplifications, it becomes possible to explicitly calculate the terms that appear in the second-order adiabatic-response approximation \eqref{eq: rho ad resp}. 
Specifically, the state of the quantum subsystem is given by 
\begin{equation}
\label{eq: mf 2nd order}
    \brho_{t}=\bar{\varrho}^{\qu}_{t} + \brho^{(x)}_{t}p_t + \brho^{(2x)}_{t}p^{2}_t + \brho^{(p)}_{t}\dot{p}_t,
\end{equation}
where
\begin{equation}
\label{eq: mf state exp}
    \brho^{(x)}_x = \frac{1}{Nm}\Linmf{x}^{-1} \bar{\varrho}_{x}^{\qu}{'}, \qquad
    \brho^{(2x)}_{x} =\frac{1}{N^2m}\Linmf{x}^{-1}\brho_{x}^{(x)}{'}, \qquad
    \brho^{(p)}_x =\Linmf{x}^{-1}\brho^{(x)}_{x}
\end{equation}
and the derivative of the MF steady state is
\begin{equation}
\label{eq: mf dx rho}
    \bar{\varrho}_{x}^{\qu}{'}= \ \frac{\chimf_{x}^{-}}{\chimf_{x}^{+}}\delta \bar{F}_{x}^{\hbar\omegaa}\bar{\varrho}_{x}^{\qu}\partial_{x}\bigg(\frac{\chimf_{x}^{+}}{\chimf_{x}^{-}}\bigg),
\end{equation}
with $\delta \bar{F}_{x}^{\hbar\omegaa}=-\partial_{\hbar\omegaa}H_{x}^{\qu}+\text{Tr}\big[\partial_{\hbar\omegaa}H_{x}^{\qu}\bar{\varrho}_{\qu}\big]$.

\newcommand{\KinHS}[1]{\mathsf{K}^{\ddagger}\big[#1\big]}

\subsubsection{Quantum coupling force}

The terms that appear in the expression \eqref{eq: Ryd qf} for the quantum force can now be determined in the MF picture. 
The deterministic force contribution is given by
\begin{equation}
\label{eq: mf F_x}
    \bar{\mathcal{F}}_{x}'/N=\text{Tr}\big[{H_{x}^{\qu}}'\bar{\varrho}_{x}^{\qu}\big]/N=V_{x}'\bigg(\frac{\bar{\chi}_{x}^{+}}{\Xmf_{x}}\bigg)^{2}.
\end{equation}
This expression is independent of $N$ as expected since in the rescaled picture the cavity length increases with the number of atoms and the relative interatomic spacing is unchanged.
The systematic force that the quantum system exerts on the mirror is therefore independent of the number of atoms in the chain.

The quantum diffusion constant \eqref{eq: Dqu def} involves the adjoint Lindblad super-operator $\mathsf{L}^\ddagger_x$ and, in the MF picture, becomes
\begin{equation}
    \bar{D}_{x}^{\qu}=-\llangle\big(\Linmf{x}^{\ddagger}\big)^{-1}\delta \bar{F}_{x};\delta \bar{F}_{x}\rrangle_{\bar{\varrho}^{\qu}},
\end{equation}
where $\delta \bar{F}_{x}=F_{x}+\bar{\mathcal{F}}_{x}'$ and
\begin{equation}
   \Linmf{x}^{\ddagger}\ \circ\equiv\Hin{x}\circ+ \sumN{k}{N}\Bigl\{\bar{\chi}^{+}\KinHS{\sigma_{k}^{+}}\circ+\bar{\chi}^{-}\KinHS{\sigma_{k}^{-}}\circ\Bigr\}
\end{equation}
with
\begin{equation}
    \KinHS{Y}\circ =  Y^{\dagger}\circ Y - \frac{1}{2} Y^{\dagger}Y\circ - \frac{1}{2}
    \circ Y^{\dagger}Y.
\end{equation}
This expression can be evaluated explicitly, which yields the result
\begin{equation}
    \bar{D}_{x}^{\qu}={V_{x}'}^{2}\frac{(\chimf_{x}^{+})^2\chimf_{x}^{-}\big(8\chimf_{x}^{+}+\chimf_{x}^{-}\big)}{2N\Xmf_{x}^5},
\end{equation}
for details see App.~\ref{App: Dqu}. We recall that the quantum diffusion constant is related to the MF quantum noise through the relation
\begin{equation}
    \langle \bar{\xi}_{x,t}^{\qu}\bar{\xi}_{x,t'}^{\qu}\rangle=2\bar{D}_{x}^{\qu}\delta_{t-t'}.
\end{equation}
Thus the Langevin equations that describe the dynamics of the effective classical degree of freedom become
\begin{subequations}
\label{eq: mf Lange}
\begin{align}
    \dot{x}_t&=\ p_t/m,
    \\
    \dot{p}_t&= -c(x_t-x_0)-\bar{\mathcal{F}}_{t}'/N^2-(\betatx{qu}\bar{D}_{t}^{\qu}+\betatx{cl}D^{\cl})\dot{x}_t +\xi_{t}^{\cl}/N+\bar{\xi}_{t}^{\qu}/N,
\end{align}
\end{subequations}
where all quantities are known explicitly and $\bar{\mathcal{F}}'$ and $\bar{D}^{\qu}$ depend only on time through $x_t$. 
We therefore have derived a full dynamical description of our quantum-classical engine in the MF picture.

\subsection{Thermodynamic processes}

\subsubsection{Setting}

We are now ready to explore the stochastic thermodynamics of our quantum-classical hybrid machine.
We first observe that, within the mean-field picture, the rates of work applied to the classical load, heat exchange with the quantum working system and heat uptake from the quantum reservoir, per particle, are given by
\begin{subequations}
\label{eq:dot energy def}
    \begin{align}
        \dot{w}_t&=V_{t}'\bigg(\frac{\chimf_{t}^{+}}{\Xmf_{t}}\bigg)^{2}\dot{x}_t
        \\
        \dot{q}^{\text{ex}}_t&=\betatx{\qu}{V_{t}'}^{2}\frac{(\chimf_t^{+})^2\chimf_t^{-}\big(8\chimf_t^{+}+\chimf_t^{-}\big)}{2N\Xmf_t^5}\dot{x}_t^2,
        \\
        \dot{q}^{\text{d}}_t&=\text{Tr}\big[H_{t}^{\qu}\Linmf{t}\brho^{(x)}_{t}\big]p_{t}+\text{Tr}\big[H_{t}^{\qu}\Linmf{t}\brho^{(p)}_{t}\big]\dot{p}_{t}
       +\text{Tr}\big[H_{t}^{\qu}\Linmf{t}\brho^{(2x)}_{t}\big]p^{2}_{t}\label{eq: mf qd},
    \end{align}
\end{subequations}
respectively, where we provide the explicit result for $\dot{q}^{\text{d}}_t$ in App.~\ref{App: q terms}. 
These expressions are obtained by inserting the mean-field approximations \eqref{eq: mf 2nd order}-\eqref{eq: mf F_x} into the general formulas \eqref{eq: dot qex}. 
As we will show in the following, they make it possible to infer the full distributions of thermodynamic quantities on the level of single trajectories from numerical simulations of the Langevin equations \eqref{eq: mf Lange}.

We consider two examples of thermodynamic processes: a single stroke expansion of the cavity and a two-stroke engine cycle. The Langevin dynamics described by Eq.~\eqref{eq: mf Lange}, is simulated using an Euler-Maruyama method, which we explain further in App.~\ref{App: numerics}. 
Throughout this section we consider a chain with $N=100$ Rydberg atoms. 
The interaction strength between nearest-neighbour atoms is $V_x=v/x^3$ where we choose $v=5\hbar\omegaa x_0^3$ as is typical for the baseline interaction strength of trapped Rydberg atoms \cite{Hoening2014}. 
We also fix the spring constant, the classical diffusion constant and the inverse temperature of the classical reservoir at $c=\hbar\omegaa/x_0^2$, $D^{\cl}=10^{5}\hbar^{2}\omegaa/x_0$ and $\betatx{cl}=0.1/\hbar\omegaa$. 
Since all relevant thermodynamic quantities are linear in $N$ within the mean-field picture, we will only consider work and heat contributions per atom in the chain.

\subsubsection{Single-stroke expansion}

\begin{figure}[t]
    \centering
    \includegraphics[width=\textwidth]{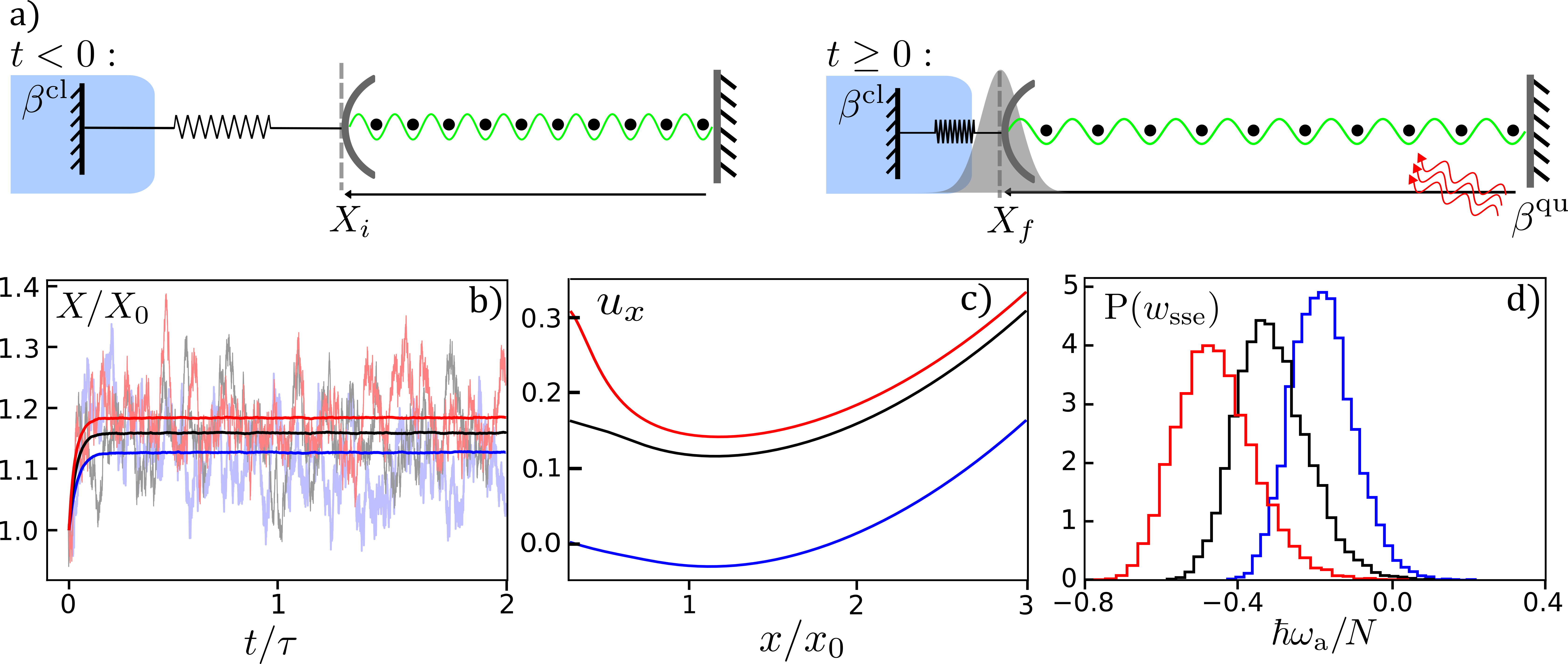}
    \caption{
    	a) Schematic outline of a single-stroke expansion. 
    	For $t<0$, the movable mirror is held at some initial position $X_{i}$. 
    	At $t=0$, the mirror is released and the cavity expands under the internal radiation
    	pressure, generated by the quantum system, which is coupled to a noisy laser playing 
    	the role of a thermal bath with inverse temperature $\betatx{qu}$. 
    	b) Plots of the average mirror position and individual trajectories over time, shown 
    	for three different inverse temperatures of the quantum bath,
    	$\betatx{qu}=\betatx{cl}$ (blue), $0.5\betatx{cl}$ (black) and $0.1\betatx{cl}$ (red). 
    	The time scale over which the expansion takes place is $\tau=2\times 10^{6}\omegaa^{-1}$.
    	c) The rescaled effective potential $u_x=\frac{1}{2}c(x-x_0)^{2}+\mathcal{F}_x/N^{2}$
    	acting on the classical degree of freedom for the same temperatures as in b). 
    	d) Histograms of the work distribution for $10^4$ realizations of the same 
    	single-stroke expansions in b).}
    \label{fig: ITE plot}
\end{figure}

The single-stroke expansion is illustrated in Fig.~\ref{fig: ITE plot}a). 
We consider the movable mirror to be fixed at an initial position $X_{i}\equiv X_{0}$, equal to its natural resting position in the absence of any radiation pressure inside the cavity, i.e., for $\betatx{qu}=\infty$ and $\bar{\mathcal{F}}'_{x}=0$. 
Upon turning on the noisy laser, the atoms in the chain are exposed to an effective environment with inverse temperature $\beta^{\qu}>0$ which leads to a rise in radiation pressure and thus a non-zero quantum coupling force. 
As a result, once the mirror is released, the cavity expands and the mirror settles, on average, to a new equilibrium position $X_{f}$.
We plot the average position of the mirror together with exemplary single trajectories in Fig.~\ref{fig: ITE plot}b). 
The larger the effective temperature of the quantum reservoir, the further the mirror is pushed by the cavity and the more its final equilibrium position $X_f$ deviates from its initial position $X_0$. 
The total work done by the quantum system on the movable mirror during the expansion is given by 
\begin{equation}
    w_{\text{sse}}=\int_0^{t} dt \ \dot{w}_t
\end{equation}
where $t$ has to be chosen sufficiently large such that the mirror has reached its new equilibrium position. 
Fig.~\ref{fig: ITE plot}b) indicates that no systematic motion of the mirror occurs anymore after $t=10^{6}\omegaa^{-1}$; we therefore set $t=2\times10^{6}\omegaa^{-1}$ in the following. 
In part c) of Fig.~\ref{fig: ITE plot} we plot the rescaled potential $u_x=c(x-x_0)^{2}/2+\mathcal{F}_x/N^2$, which is responsible for the average final resting position of the mirror.
Since we obtain the time-resolved position $x_t$ of the mirror from our simulation of the Langevin equation \eqref{eq: mf Lange}, we can now calculate $w_{\text{sse}}$ for any individual trajectory.
Upon repeating this simulation sufficiently many times, we can determine the distribution of this quantity, which is given by
\begin{equation}
    \mathrm{P}(w_{\text{sse}})=\frac{1}{M}\sum_{m=1}^{M}\Pi_\epsilon [w_{\text{sse}}-w_{\text{sse}}^{(m)}],
\end{equation}
for sufficiently large $M$, where $w_{\text{sse}}^{(m)}$ is the work performed along the $m$-th trajectory and 
\begin{equation}
    \Pi_{\epsilon}[x]=\begin{cases}
        &1\quad\text{if}\quad-\epsilon/2\leq x\leq\epsilon/2 \\
        &0\quad\text{otherwise}
    \end{cases}
\end{equation}
is a boxcar function with width $\epsilon$. 
The distributions are plotted in part d) of Fig.~\ref{fig: ITE plot}. 
These plots show nearly Gaussian distributions with a slight asymmetry. 
Their mean value depends on the inverse temperature of the quantum bath, which determines the radiation pressure inside the cavity. 
In addition, we observe that the width of the work distributions increases with the temperature of the quantum reservoir, due to enhanced thermal fluctuations.

\subsubsection{Two-stroke engine cycle}

\begin{figure}[t]
    \centering
    \includegraphics[width=\textwidth]{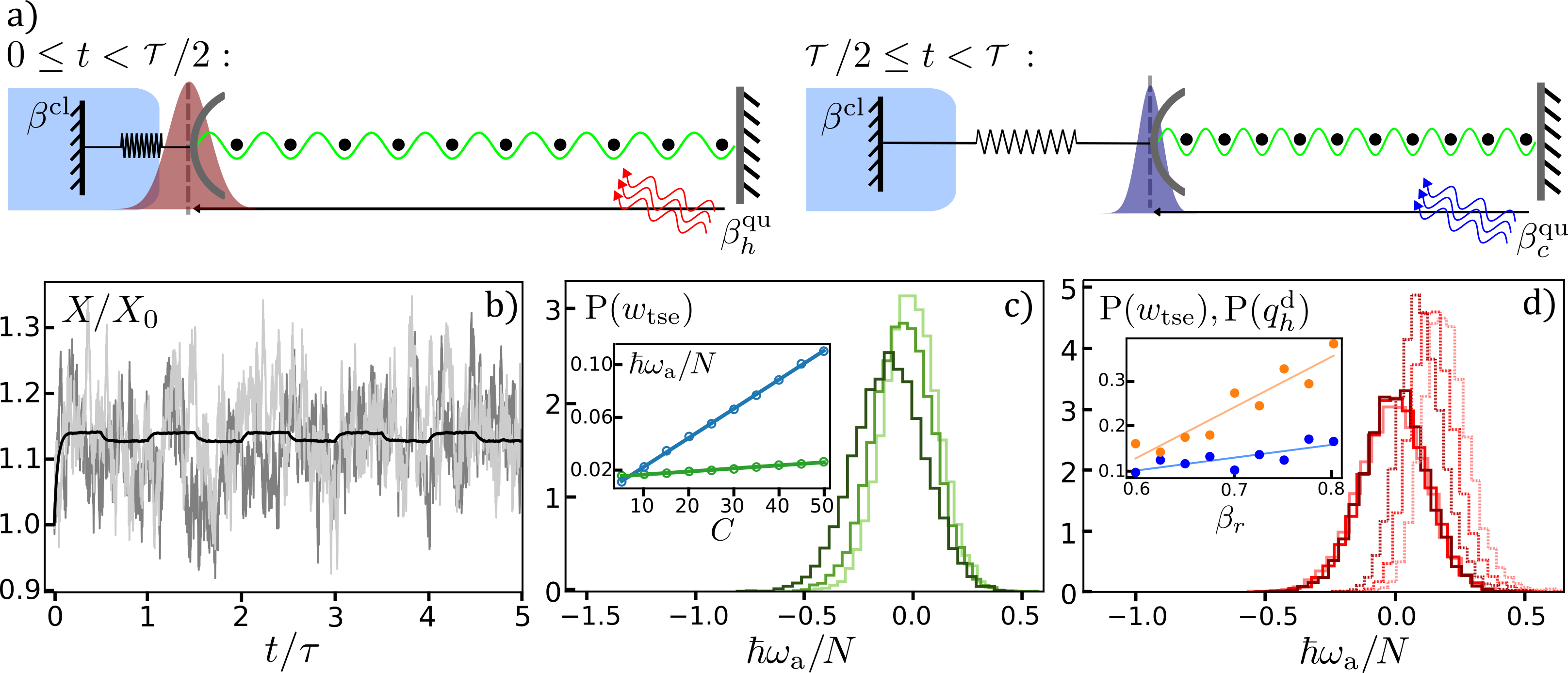}
    \caption{
    	a) Schematic outline of a two-stroke cycle, driven by periodic temperature modulations. 
    	b) Average position of the mirror taken over $10^4$ trajectories for 5 cycles (black solid)
    	with $\betatx{qu}_h=0.8\betatx{cl}$ and $\betatx{qu}_c=\betatx{cl}$, 
    	$\tau=2\times 10^{6}\omegaa^{-1}$ fixed throughout all plots. 
    	The shaded lines show the position of the mirror from two individual trajectories. 
    	c) Plots of the work distribution over $C=5$ (light green), $20$ (green) and $50$ 
    	(dark green) cycles, with $\betatx{qu}_h=0.8\betatx{cl}$. 
    	The inset shows the mean $W$ (blue circles) and variance $\text{Var}(W)$ (green circles) 
    	of the output work as a function of the number of cycles, ranging from $C=5-50$, with 
    	solid lines showing linear fits. 
    	d) Distributions of output work $w_{\text{tse}}$ (solid) and input heat $q^{\text{d}}_h$
    	(dotted) for $10^4$ realizations of a single cycle, for different values of 
    	$\beta_r=\betatx{qu}_{h}/\betatx{qu}_c$. 
    	Here $\betatx{qu}_h$ is changed such that $\beta_r=0.7$ (brown), $0.75$ (red) and 
    	$0.8$ (pink). 
    	The inset shows the corresponding efficiency $\eta$ (blue) in units of  
    	$\eta_{\text{C}}$ as well as the rescaled relative fluctuation of the output work, 
    	$\sqrt{\text{Var}(W)/W^2}/100$ (orange), for a range of temperature ratios $\beta_r$; 
    	solid lines show linear fits.}
    \label{fig: two-stroke cycle}
\end{figure}

As our second example, we consider a two-stroke engine cycle with the quantum system acting as the working medium. 
The engine cycle is realized by periodic modulation of the external laser field, so that the effective inverse temperature of the quantum reservoir oscillates between the two values $\betatx{qu}_c\equiv\betatx{cl}$ and $\betatx{qu}_h<\betatx{qu}_c$. 
The process is outlined in Fig.~\ref{fig: two-stroke cycle}a. 
In the first stroke, where $0\leq t<\tau/2$, the cavity expands under the increased radiation pressure induced by the quantum reservoir at $\betatx{qu}_h$. In the second stroke, where $\tau/2\leq t<\tau$, the laser intensity is adjusted to emulate the cold quantum bath with $\betatx{qu}_c$, so that the movable mirror of the cavity relaxes to a new position. 
The period of the cycle is $\tau=2\times10^{6}\omegaa^{-1}$ so that the adiabatic-response assumption is justified within the individual strokes, except for minor corrections incurred right after the switching of the temperature. 
That is, the quantum system quickly relaxes to its new adiabatic-response state after the external field changes at the start of each stroke. 
Here the temperature difference between the hot and cold stroke must also be small relative to the internal energy scale of the quantum system in order for the thermodynamics of a two-stroke cycle to be consistently described by the adiabatic response framework without introducing a systematic expansion in $\Delta\betatx{qu}=\betatx{qu}_h-\betatx{qu}_c$ \cite{Eglinton2022}. 

We simulate the position of the movable mirror over $5$ cycles, and average over $10^4$ trajectories in Fig.~\ref{fig: two-stroke cycle}b). 
We find that the mirror settles into a periodic steady state, within approximately two cycles. 
Once the engine has relaxed into a periodic steady state we can calculate its accumulated net output over $C$ cycles,
\begin{equation}
    w_{\text{tse}}(C)=\int_t^{t+C\tau} dt' \ \dot{w}_{t'}.
\end{equation}
In part c) of Fig.~\ref{fig: two-stroke cycle} we show three plots for the work distributions for $C=5$, $20$ and $50$ cycles. 
The inset shows that both the mean value and variance of these distributions grow linearly in $C$, as one would expect. 
As a second quantity of interest, we consider the heat uptake from the effective quantum reservoir in one cycle $q^{\text{d}}_h$, for different temperature ratios, $\beta_{r}=\betatx{qu}_h/\betatx{qu}_c$. 
Following the stochastic expression in Eq.~\eqref{eq: qh,qc}, the heat uptake in the hot stroke is given by
\begin{equation}
    q^{\text{d}}_h=\int_t^{t+\tau} dt' \ \theta^h_{t'}\dot{q}_{t'}^{\text{d}}+\varepsilon_{x,t_h},
\end{equation}
where $\theta^h_t$ is 1 during the hot phase of the cycle and zero otherwise and $t_h$ is the time at which the quantum reservoir switches from cold to hot.
For our Rydberg engine we take $\dot{q}^{d}_{t}$ from Eq.~\eqref{eq: mf qd} and the contribution from switching the temperature at the start of the stroke is
\begin{equation}
    \varepsilon_{x,t_h}=\lim_{\epsilon\rightarrow0}\biggl(\frac{\chimf^{+}_{t_h+\epsilon}}{\chimf_{t_h+\epsilon}}\Big(\hbar\omegaa+V_{t_h+\epsilon}\frac{\chimf^{+}_{t_h+\epsilon}}{\chimf_{t_h+\epsilon}}\Big)
    -\frac{\chimf^{+}_{t_h-\epsilon}}{\chimf_{t_h-\epsilon}}\Big(\hbar\omegaa + V_{t_h-\epsilon}\frac{\chimf^{+}_{t_h-\epsilon}}{\chimf_{t_h-\epsilon}}\Big)\biggr)
\end{equation}
in the MF picture. 
The corresponding distributions are shown in part d) of Fig.~\ref{fig: two-stroke cycle} together with the work distributions for $\beta_r=0.7, 0.75$ and $0.8$. 
Finally, from the mean values $W$ and $Q^{\text{d}}_h$ of the output and input distributions, we can calculate the efficiency of our engine $\eta=-W/Q^{\text{d}}_h$, which is plotted in the inset of part d) of Fig.~\ref{fig: two-stroke cycle} in units of the Carnot value $\eta_{\text{C}} = 1- \betatx{qu}_h/\betatx{qu}_c$, along with the relative fluctuations of the output work which characterise the stability of the output \cite{Holubec2014}, that is, the constancy of the engine \cite{Pietzonka2018}. 
We find that the efficiency of our engine reaches up to $\sim 18\%$ of the Carnot value. 
This figure is plausible given that we have not systematically optimized the parameters and driving protocol of our model. 
We leave this task to future research. 
The relative work fluctuations are seen to increase with $\beta_r$. This behaviour primarily results from the fact that the mean output work decreases with $\beta_r$ as the mirror's equilibrium positions in the hot and the cold stroke move closer together. 
In summary, our results demonstrate that the general framework developed in this article makes it possible to obtain the full statistics of thermodynamic processes at the quantum-classical boundary without the necessity of invoking additional measurement protocols, which was our main objective here. 

\section{Perspectives}\label{Sec:Perspectives}

Our case study demonstrates that the essential characteristics of microscopic thermal machines can be realized in quantum-classical hybrid systems. 
In particular, we have seen that a classical degree of freedom with an observable trajectory provides both a physically intuitive means of work extraction and a powerful tool to infer the statistics of quantum thermodynamic processes.
It now remains to find suitable experimental platforms to realize such quantum-classical hybrid engines. 
In our analysis, we have rescaled the cavity length with $x_{0}=5\mu $m, which corresponds to the typical inter-particle spacing for trapped Rydberg atoms that can be achieved in current experiments \cite{Muller2008}. 
The magnitudes of the dipole-dipole interactions and the collective decay rates were chosen as $v/x_0^3=5\hbar\omegaa$ and $\kappa=10^{-2}/\hbar\omegaa^2$, where the energy difference between the ground and the Rydberg state was set to the realistic value $\omegaa=1 $MHz \cite{Hoening2014,Muller2008,Higgins2017,Higgins2017b}. 
The mass of the classical degree of freedom scales as $m\propto\hbar/x_0^2\omegaa$, which for our model would lead to a value of the order of $10^{-20}g$. Such a tiny mass cannot be plausibly associated with a nano-mechanical resonator such as a movable mirror. 
It may, however, be achieved with cold-atoms based opto-mechanical systems, where the mechanical degrees of freedom are formed by collective vibration modes of atomic ensembles \cite{Schleier2011}. 
A second avenue towards a more feasible setting could be to decrease the typical inter-particle distance $x_0$, which would lead to an increase in the radiation pressure inside the cavity and thus the force acting on the mirror, as is reflected by the scaling of the effective mass with $x_0^{-2}$. 
Overall, our model is meant to conceptualize and illustrate the basic idea and potential benefits of quantum-classical hybrid machines and we leave it to future research to find truly realistic ways of constructing such devices.

On the conceptual side, our framework provides a systematic step towards bridging the long-standing gap between classical stochastic and quantum thermodynamics. 
Since the concepts of the former theory can be directly applied to the classical degree of freedom, our results lay the ground work for a variety of future research directions, such as the potential formulation of fluctuation theorems and thermodynamic uncertainty relations for quantum-classical hybrid systems or the search for universal trade-off relations between the figures of merit of hybrid thermal machines. 
The model that we have developed for such quantum-classical thermal machines may be regarded as a semi-classical limit of so-called quantum autonomous machines, which are driven only by heat fluxes and do not rely on external control mechanisms to perform useful tasks \cite{Mitchison2019a}. 
Instead, the output of such devices can be channeled directly into additional degrees of freedom, which are akin to classical flywheels and couple directly to the working system. 
These storage systems are usually described quantum-mechanically, along with the working system. 
It would be interesting to explore whether and how thermodynamically consistent quantum-classical hybrid models can indeed be obtained from such quantum autonomous settings in certain limiting scenarios, where the storage degrees of freedom behave effectively classically.
In a wider perspective, it would be desirable to extend our approach beyond the limits of the adiabatic-response regime and ultimately to derive a thermodynamically consistent theory of quantum-classical systems from fundamental principles. 
Useful lessons to this end could be drawn, for instance, from physical chemistry, where hybrid models have long been used to simplify fully quantum descriptions of complex molecular systems \cite{Kapral2015,Kappral2016}, from field theory, where the lack of a quantum theory of gravity has long been a driving force for research on quantum-classical dynamics \cite{Kuo1993,Anderson1995}, or from mathematical physics, where stochastic descriptions of quantum-classical systems are currently receiving renewed attention \cite{Diosi2014,Oppenheim2022,Oppenheim2023,Layton2023,Lajos2023}. 
Our thermodynamics-based approach adds a further perspective to this research and might, owing to the universality of thermodynamics as a physical theory, eventually help to unify existing and forthcoming results across different fields. 

\section*{Acknowledgments}
JE acknowledges support from the German Academic Exchange Service (DAAD).
This work was supported by the Medical Research Council (Grants No. MR/S034714/1 and MR/Y003845/1) and the Engineering and Physical Sciences Research Council (Grant No. EP/V031201/1). KB acknowledges insightful discussions with Keiji Saito. 
FC is indebted to the Baden-W\"urttemberg Stiftung for the financial support by the Eliteprogramme for Postdocs.
IL and FC acknowledge funding from the European Union’s Horizon Europe research and innovation program under Grant Agreement No. 101046968 (BRISQ). They also are grateful for financial support through the Deutsche Forschungsgemeinschaft (DFG, German Research Foundation) through the Research Unit FOR 5413/1, Grant No. 465199066.
This work was supported by the University of Nottingham and the University of T\"{u}bingen’s funding as part of the Excellence Strategy of the German Federal and State Governments, in close collaboration with the University of Nottingham.

\section*{Data access statement}
All plotted data were generated from the equations and simulations described in the main text and appendix. 
No further data were created by the research presented in this article. 

\appendix 

\newcommand{\sigz}[1]{\sigma^{z}_{#1}}

\section{Mean-field theory}

\subsection{\label{App: Dqu}Quantum diffusion constant}

We recall the expression
\begin{equation}
    D_{x}^{\qu} = \int_0^{\infty} dt \ \llangle \delta \hat{F}_{x,t} ;\delta \hat{F}_{x,0}\rrangle_{\varrho^{\qu}}
\end{equation}
for the quantum diffusion constant, which was defined in Eq.~\eqref{eq: Dqu def}. 
In this appendix, we first show how this quantity can be obtained analytically within the mean-field picture introduced in Sec.~\ref{sec:Rydb eng}, and then compare the resulting expression with numerically exact calculations for a small number $N$ of Rydberg atoms.

Within the mean-field approximation, the quantum diffusion constant is given by
\begin{equation}
\label{app: mfDq}
    \bar{D}^{\qu}_x=\int_0^{\infty} dt \ \llangle \delta \hat{\bar{F}}_{x,t} ;\delta \hat{\bar{F}}_{x,0}\rrangle_{\varrho^{\qu}}
\end{equation}
where $\delta \hat{\bar{F}}_{x,t}$ denotes the shifted quantum force operator in the Heisenberg picture with respect to the adjoint mean-field generator
\begin{equation}
\label{app: mf gen}
    \Linmf{x}^{\ddagger}\circ=\Hin{x}\circ+ \sumN{k}{N}\Bigl\{\bar{\chi}^{+}\KinHS{\sigma_{k}^{+}}\circ+\bar{\chi}^{-}\KinHS{\sigma_{k}^{-}}\circ\Bigr\}
\end{equation}
with
\begin{equation}
    \Hin{x}=-\frac{i}{\hbar}\big[\circ,H^{\qu}_x\big]\quad\text{and}\quad\KinHS{Y}\circ =  Y^{\dagger}\circ Y - \frac{1}{2} Y^{\dagger}Y\circ - \frac{1}{2} \circ Y^{\dagger}Y.
\end{equation}
This object satisfies the differential equation
\begin{equation}
\label{app: ddt deltaF}
\frac{d}{dt}\delta\bar{F}_{x,t}=\Linmf{x}^{\ddagger}\delta \bar{F}_{x,t}    
\end{equation}
within respect to the initial condition
\begin{equation}
\label{app: mf F}
    \delta\bar{F}_{x,0}=F_x+\bar{\mathcal{F}}'_{x,t}=-V_x'\biggl(\sum_{k=1}^{N}n_kn_{k+1}-\text{Tr}\biggl[\sumN{k}{N}n_k n_{k+1}\bar{\varrho}_{x}^{\qu}\biggr]\biggr)/N,
\end{equation}
where $F_x$ is the effective force operator defined in Eq.~\eqref{eq: F=-H} and $\bar{\mathcal{F}}_x$ denotes the equilibrium free energy of the quantum system, defined in Eq.~\eqref{eq: mf F_x}.

This initial value problem can be solved by exploiting the fact that $\delta \bar{F}_x$ belongs to an invariant subspace of $\Linmf{x}^{\ddagger}$. 
To this end, we define a sequence of operators
\begin{equation}
    \Linmf{x}^{\ddagger}\delta \bar{F}^{1}_{x}=\delta \bar{F}_{x}^{2},\qquad
    \Linmf{x}^{\ddagger}\delta \bar{F}_{x}^{2}=\delta \bar{F}_{x}^3,\qquad\dots,\qquad
    \Linmf{x}^{\ddagger}\delta \bar{F}_{x}^{N}=\delta \bar{F}_{x}^{N+1}.
\end{equation}
If $\delta\bar{F}^{N+1}_x$ can be expressed as a linear combination of $\delta\bar{F}_x^{1},\dots,\delta\bar{F}^{N}_x$, i.e., if there exists constants $c_x^k$ so that $\delta\bar{F}^{N+1}_x=\sum_{k=1}c_x^k\delta \bar{F}^{k}_x$, the solution of Eq.~\eqref{app: ddt deltaF} can be written as
\begin{equation}
\label{app: Ldf with A}
    e^{\Linmf{x}^{\ddagger}}\delta \bar{F}_{x}=(1,0,\dots,0)\cdot e^{\mathbf{A}_x s}\delta \bar{\mathbf{F}}_{x},
\end{equation}
where $\delta\bar{\mathbf{F}}_x= \big(\delta \bar{F}_{x}^{1}, \delta \bar{F}_{x}^{2}\dots,\delta \bar{F}_{x}^{N}\big)^{\text{T}}$ and the matrix $\mathbf{A}_x$ is given by
\begin{equation}
    \Linmf{x}^{\ddagger}\delta\bar{\mathbf{F}}_x=\mathbf{A}_x\delta\bar{\mathbf{F}}_x.
\end{equation}
Upon using the expressions \eqref{app: mf gen} and \eqref{app: mf F}, we find
\begin{subequations}
\begin{align}
    \Linmf{x}^{\ddagger}\delta \bar{F}_{x}^1 &= -\frac{V_x'}{N}\sumN{k,l}{N}\Bigl\{\bar{\chi}_{x}^{+}\KinHS{\sigma^+_k}n_{l}n_{l+1} +\bar{\chi}_{x}^{-}\KinHS{\sigma^-_k}n_{l}n_{l+1}\Bigr\}\\
    &=-\frac{V_x'}{N}\biggl(-2\chimf^-\sumN{l}{N}n_l n_{l+1}+\chimf^+\sumN{l}{N}\Bigl\{n_l g_{l+1}+g_l n_{l+1}\Bigr\}\biggr) \equiv \delta \bar{F}_{x}^{2},  \nonumber \\
    \Linmf{x}^{\ddagger}\delta \bar{F}_{x}^2 &=\frac{V_x'}{N}\sumN{k,l}{N}\bigg\{\Big(\chimf_x^{+}\KinHS{\sigma^+_k}+\chimf_x^{-}\KinHS{\sigma^-_k}\Big)\Big(2\chimf^-n_l n_{l+1}-\chimf^+\big(n_l g_{l+1}+g_l n_{l+1}\big)\Big)\bigg\} \nonumber \\
    &=-(3\chimf_{x}^{-} + \chimf_{x}^{+})\delta \bar{F}_{x}^{1}+\frac{2V_x'}{N}\bigg((\chimf_{x}^{-})^2\sumN{l}{N}n_l n_{l+1} - (\chimf_{x}^{+})^2\sumN{l}{N}g_l g_{l+1}\bigg)
    \equiv\delta \bar{F}_{x}^{3},
    \\
    \Linmf{x}^{\ddagger}\delta \bar{F}_{x}^{3}&= -3\Xmf\delta \bar{F}_{x}^2 - 2\Xmf_{x}^{2}\delta \bar{F}_{x}^{3},
\end{align}
\end{subequations}
with the coefficient matrix
\begin{equation}
    \mathbf{A}_{x}=\begin{pmatrix}
        0 & 1 & 0 \\
        0 & 0 & 1 \\
        0 & -3\Xmf_{x} & -2\Xmf_{x}^{2}
    \end{pmatrix}.
\end{equation}
This matrix can be easily exponentiated, which, upon insertion into the Eqs.~\eqref{app: Ldf with A} and \eqref{app: mfDq} yields the compact expression
\begin{equation}
    \bar{D}^{\qu}_x={V_{x}'}^{2}\frac{(\chimf_x^{+})^2\chimf_x^{-}\big(8\chimf_x^{+}+\chimf_x^{-}\big)}{2N\Xmf_x^{5}},
\end{equation}
for the quantum diffusion constant in the mean-field approximation.

\begin{figure}[t]
    \centering
    \includegraphics[width=.6\textwidth]{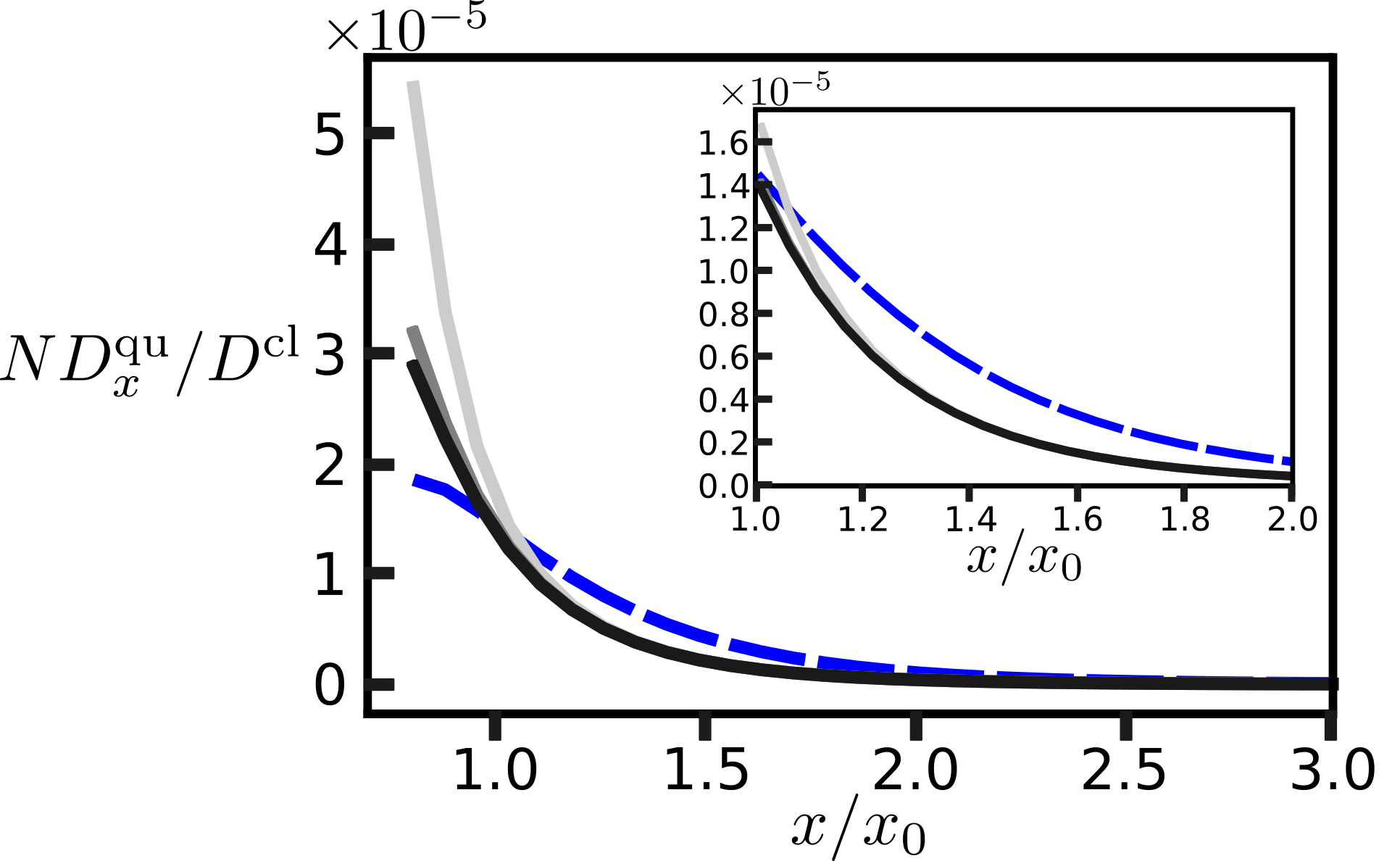}
    \caption{
    	Plot of MF and exact results for the quantum diffusion constant $D_{x}^{\qu}$ as a 
    	function of the normalized inter-atomic distance $x/x_0$ for the inverse temperature
    	$\betatx{qu}=0.1/\hbar\omegaa$. 
    	In the relevant regime, the behaviour of the MF result (blue dashed) is in good agreement
    	with the exact plots, which are numerically determined for $N=5$ (light grey),
    	$N=7$ (grey) and $N=9$ (black). 
    	The inset highlights the results around the region where the typical trajectories of the
    	two-stroke cycle occur in Figs.~\ref{fig: ITE plot} and \ref{fig: two-stroke cycle}.}
    \label{fig: MF valid plot}
\end{figure}

This result provides a useful benchmark for the quality of the mean-field approximation. 
For a quantitative comparison, we numerically solve the exact equation of motion, $\delta F_{x,t}=\Lin{x}^{\ddagger}\delta F_{x,t}$, for the shifted force operator in the Heisenberg picture and then evaluate $D^{\qu}_x$ with the help of Eq.~\eqref{eq: Dqu def}. 
The results of this calculation are plotted in Fig.~\ref{fig: MF valid plot} for a relevant range of system parameters and Rydberg chains with $N=5,7,9$ atoms. 
We find that, within the spatial region where the mirror settles during the two-stroke cycle described in Sec.~\ref{sec:Rydb eng}, cf. Fig.~\ref{fig: two-stroke cycle}, the mean-field and exact results are indeed in good agreement up to smaller deviations. 
In the limit $x\rightarrow 0$, the interaction potential $V_x$ diverges and the mean-field approximation breaks down.
However, within the situations discussed in the main text, $x$ is effectively limited to a finite range around $x_0$, where the interaction strength is still sufficiently weak for the mean-field approximation to be applicable.

\label{App: mf plot}

\subsection{Stochastic dissipated heat current in the mean-field picture}

\label{App: q terms}

The three terms that appear in the expression \eqref{eq: mf qd} for the stochastic dissipated heat current within the mean-field approximation are given by
\begin{subequations}
\label{Apeq: qd terms}
    \begin{align}
        \text{Tr}\big[H_{x}^{\qu}\Linmf{x}\brho^{(x)}_{x}\big]&=\frac{1}{Nm}\text{Tr}\big[H_{x}^{\qu}{\bar{\varrho}^{\qu\prime}_{x}}\big]
        \\
        &=\frac{1}{m}\ \frac{\chimf_{x}^{-}}{\chimf_{x}}\Biggl(\biggl(\omegaa+2\frac{\chimf^+_x}{\chimf_x}V_x\biggr)\biggl(1-\frac{\chimf^+_x}{\chimf_x}\biggr)\Biggr)\partial_{x}\bigg(\frac{\chimf_{x}^{+}}{\chimf_{x}^{-}}\bigg),\nonumber
    \\
    \text{Tr}\big[H_{x}^{\qu}\Linmf{x}\brho^{(p)}_{x}\big]&=\frac{1}{N^2m}\text{Tr}\big[H_{x}^{\qu}\Linmf{x}^{-1}{\bar{\varrho}_{x}^{\qu}}{'}\big]=\int_0^\infty ds \ \text{Tr}\big[H^{\qu}_{x} e^{\Linmf{x}^{\ddagger}s}\bar{\varrho}^{\qu\prime}_x\big]
    \\
    &=-\frac{(\chimf_x^{-})^2\big(2V_{x}\chimf_x^{+} + \omegaa\Xmf_x\big)}{Nm\Xmf_x^{4}}\partial_{x}\bigg(\frac{\chimf_x^{+}}{\chimf_x^{-}}\bigg),\nonumber
    \\    \text{Tr}\big[H_{x}^{\qu}\Linmf{x}\brho^{(2x)}_{x}\big]&=\frac{1}{Nm^2}\text{Tr}\big[H_{x}^{\qu}\partial_x\big(\Linmf{x}^{-1}{\bar{\varrho}^{\qu\prime}_{x}}\big)\big]
    \label{appeq: 9c}\\
    &=-\partial_{x}\Bigg(\frac{(\chimf_x^{-})^2\big(2V_{x}'\chimf_x^{+} + \omegaa\Xmf_x\big)}{m^2\Xmf_x^{4}}\partial_{x}\bigg(\frac{\chimf_x^{+}}{\chimf_x^{-}}\bigg)\Bigg) + 2V_{x}\frac{(\chimf_x^{-})^2\chimf_x^{+}}{m^2\Xmf_x^{4}}\partial_{x}\bigg(\frac{\chimf_x^{+}}{\chimf_x^{-}}\bigg).\nonumber
    \end{align}
    \label{appeq:9}
\end{subequations}
Here, the first result follows immediately by inserting the expression \eqref{eq: mf dx rho} for the derivative of the mean-field steady state. 
The second result is obtained by first noting that $\Linmf{x}\bar{\varrho}^{\qu}_x\circ=\bar{\varrho}^{\qu}_x\Linmf{x}^\ddagger\circ$ and then applying the same technique as for evaluation of the quantum-diffusion constant in the previous section. 
In the present case, only one auxiliary variable $\delta \bar{F}^{\hbar\omegaa,2}_x=\Linmf{x}^{\ddagger}\bar{F}^{\hbar\omegaa}_x$ is necessary since $(\Linmf{x}^{\ddagger})^2\bar{F}^{\hbar\omegaa}_x=-\bar{\chi}_x\delta\bar{F}^{\hbar\omegaa}$, where $\delta \bar{F}^{\hbar\omegaa}$ was defined below Eq.~\eqref{eq: mf dx rho}. 
Finally, Eq.~\eqref{appeq: 9c} follows directly by applying the chain rule
\begin{equation}
    \text{Tr}\big[H_{x}^{\qu}\partial_x\big(\Linmf{x}^{-1}{\bar{\varrho}^{\qu}_{x}}{'}\big)\big] = \partial_x \text{Tr}\big[H_{x}^{\qu}\Linmf{x}^{-1}{\bar{\varrho}^{\qu}_{x}}{'}\big] - \text{Tr}\big[{H_{x}^{\qu}}{'}\Linmf{x}^{-1}{\bar{\varrho}^{\qu}_{x}}{'}\big]
\end{equation}
and observing that
\begin{equation}
    \text{Tr}\big[{H_{x}^{\qu}}{'}\Linmf{x}^{-1}{\bar{\varrho}^{\qu}_{x}}{'}\big] = -2NV_{x}'\frac{(\chimf_x^{-})^2\chimf_x^{+}}{\Xmf_x^{4}}\partial_{x}\bigg(\frac{\chimf_x^{+}}{\chimf_x^{-}}\bigg).
\end{equation}

\section{Numerical simulations}

\label{App: numerics}

To simulate the Langevin equations \eqref{eq: mf Lange}, we implement a simple Euler-Maruyama method \cite{Gardiner1985}. 
The iterative form of the Langevin equations in discretised time is
\begin{subequations}
\begin{align}
    x_{n+1}&=x_{n} + \Delta t \ p_{n}/m, 
    \\
    p_{n+1}&=p_{n}+\Delta t \Biggl(-c\big(x_{n}-x_0\big)-\bar{\mathcal{F}}_{x_n}'/N^2 -\big(\betatx{qu} D_{x_n}^{\qu}+\betatx{\cl}D^{\cl}\big)p_{n}/m \\
    &\qquad\qquad\quad\;
    + \frac{1}{N}\biggl(\sqrt{\frac{2D^{\cl}}{\Delta t}} G_{t_n}^{\cl}+\sqrt{\frac{2D_{x_n}^{\qu}}{\Delta t}}\ G_{t_n}^{\qu}\biggr)\Biggr),\nonumber
\end{align}
\end{subequations}
where $\Delta t= t_{n+1}-t_{n}$ denotes the time step and $G_{t}\sim\mathcal{N}(0,1)$ is a random number sampled from a Gaussian distribution with mean 0 and variance 1. 
The superscript on the random variables $G^{\qu/\cl}$ indicates that the quantum or classical noise are statistically independent. 
Upon iterating these expressions over a specified time $\tau=N_{\text{steps}}\Delta t$, where $N_{\text{steps}}$ is an integer, we can simulate the dynamics of the classical degree of freedom.
At each time step, we recalculate the deterministic quantum force, the quantum diffusion constant and thus the magnitude of the Gaussian noise.

\end{document}